\documentclass[aps,prl,preprintnumbers,showpacs,nofootinbib,linenumbers]{revtex4}

\usepackage[utf8]{inputenc}
\usepackage[english]{babel}
\usepackage{amsmath}
\usepackage{amsfonts}
\usepackage{amssymb}
\usepackage{epsfig}
\usepackage{graphics,psfrag,rotating}
\usepackage{graphicx}
\usepackage{dcolumn}
\usepackage{bm}
\usepackage{float}
\usepackage{epstopdf}
\usepackage{color}
\usepackage[usenames,dvipsnames,svgnames]{xcolor}
\usepackage[colorlinks=true,
            linkcolor=red,
            urlcolor=gray,
            citecolor=blue]{hyperref}
  \usepackage{hyperref}
    \usepackage{lineno}

\newcommand{\lsim}   {\mathrel{\mathop{\kern 0pt \rlap
{\raise.2ex\hbox{$<$}}}
 \lower.9ex\hbox{\kern-.190em $\sim$}}}
\newcommand{\gsim}   {\mathrel{\mathop{\kern 0pt \rlap
{\raise.2ex\hbox{$>$}}}
\lower.9ex\hbox{\kern-.190em $\sim$}}}

\def\3nab{\tilde{\nabla}}

\def\hsp5{\hspace{5mm}}

\def\case#1/#2{\textstyle\frac{#1}{#2}}

\def\ber {\begin{eqnarray}}
\def\eer {\end{eqnarray}}
\def\bea {\begin{eqnarray}}
\def\eea {\end{eqnarray}}

\def\bc {\begin{center}}
\def\ec {\end{center}}
\def\case#1/#2{\frac{#1}{#2}}

\newcommand{\bw}{\begin{widetext}}
\newcommand{\ew}{\end{widetext}}

\newcommand{\be}{\begin{equation}}
\newcommand{\bse}{\begin{subequation}}
\newcommand{\ese}{\end{subequation}}
\newcommand{\ee}{\end{equation}}
\newcommand{\eei}{\end{eqnarray}\indent\indent}
\newcommand{\ba}{\begin{array}}
\newcommand{\ea}{\end{array}}
\newcommand{\bal}{\begin{eqnarray}}
\newcommand{\eal}{\end{eqnarray}}

\def\case#1/#2{\textstyle\frac{#1}{#2} }

\newcommand{\bver}{\begin{verbatim}}
\newcommand{\ever}{\end{verbatim}}

\begin{document}


\title{Late-Time Alleviation of the Hubble Tension from DESI DR2 BAO Measurements in CPL Cosmology with Massive Neutrinos via Bayesian Physics-Informed Neural Networks}
\author{
 Muhammad Yarahmadi$^{1}$\footnote{Email: yarahmadimohammad10@gmail.com , yarahmadi.mhd@lu.ac.ir}
}

\affiliation{Department of Physics, Lorestan University, Khoramabad, Iran}

\date{\today}

\begin{abstract}
	We present a Bayesian analysis of the Hubble constant using
	Physics-Informed Neural Networks (PINNs), applied to the standard
	wCDM model and its dynamical extension via the
	Chevallier--Polarski--Linder (CPL) parametrization, with and without
	a free summed neutrino mass $\Sigma m_\nu$. Embedding the background
	Friedmann equation into a Bayesian PINN, we reconstruct $H(z)$ in a
	data-driven, physically consistent way while propagating epistemic
	uncertainty. Combining Cosmic Chronometers, DESI DR2 BAO, and
	Pantheon+ supernovae with Planck 2018 CMB distance priors, we
	quantify the Hubble tension against Planck and SH0ES (R22). For
	wCDM, BAO-dominated combinations favor lower $H_0$, easing
	the Planck tension at the cost of a larger SH0ES discrepancy.
	Letting the equation of state evolve within CPL shifts $H_0$ upward
	and consistently favors a mildly quintessence-like
	$w_0 \gtrsim -1$ with negative $w_a$, pointing to a slowly evolving,
	non-phantom dark energy component. Adding a free $\Sigma m_\nu$
	stabilizes $H_0$ between $69.7$ and
	$71.6~\mathrm{km\,s^{-1}\,Mpc^{-1}}$, bounds
	$\Sigma m_\nu \lesssim 0.16$--$0.28~\mathrm{eV}$ ($2\sigma$), and
	reduces the SH0ES tension below $1.8\sigma$ (down to $0.83\sigma$)
	while the Planck tension persists at $1.6$--$2.8\sigma$. AIC and BIC
	both favor CPL+$\Sigma m_\nu$ despite its extra parameters, arguing
	against overfitting. A full-CMB MCMC cross-check via Cobaya, including CMB lensing, reproduces the BPINN
	posteriors -- including the neutrino-mass bound -- within 1--2$\sigma$ at a fraction of
	the cost. A mildly evolving dark energy component combined with
	sub-eV neutrino masses thus substantially eases, though does not
	fully resolve, the Hubble tension, supporting Bayesian PINNs as an
	efficient, physically consistent tool for precision cosmology beyond
	$\Lambda$CDM.
\end{abstract}


%
%


\maketitle

\section{Introduction}
\label{sec:intro}

Dark energy is the name we give to whatever is driving the observed
late-time acceleration of cosmic expansion, first seen in the
luminosity-distance measurements of high-redshift Type Ia supernovae
\cite{Riess1998,Perlmutter1999}. The simplest framework that accounts for
it is $\Lambda$CDM, where dark energy is a cosmological constant $\Lambda$
with fixed energy density and equation of state $w=-1$
\cite{Weinberg1989,Peebles2003}. $\Lambda$CDM fits a wide range of data
well, including CMB anisotropies, baryon acoustic oscillations, and
large-scale structure, as confirmed by \emph{Planck} \cite{Planck2018}.
Still, we do not know what $\Lambda$ physically is, and open problems such
as the cosmological-constant and coincidence problems keep motivating
alternatives to $\Lambda$CDM \cite{Copeland2006,Clifton2012,Bamba2012,
	Nojiri2017}.

This has led to a broad family of models in which dark energy is allowed
to evolve rather than stay fixed. Parametrizing the dark-energy equation of
state directly is a convenient, largely model-independent way to test for
such evolution. One widely used choice is the Chevallier--Polarski--Linder
(CPL) form, $w(z)=w_0+w_a\,z/(1+z)$, which interpolates smoothly between
its value today and its behavior at earlier times \cite{Chevallier2001,
	Linder2003}. CPL captures the leading-order departure from $w=-1$ while
staying well-behaved at high redshift, which is why it works well against
supernovae, BAO, and CMB distance priors \cite{Alam2004}, and why it has
become a standard benchmark for testing dynamical dark energy against
$\Lambda$CDM.

Dynamical dark-energy models like CPL are also relevant to two current
observational tensions. The first is the Hubble tension: the mismatch
between $H_0$ inferred from the CMB under $\Lambda$CDM and $H_0$ measured
directly from local distance indicators \cite{Riess2019,Verde2019}.
Letting $w$ deviate from $-1$ and evolve with redshift changes the
late-time expansion history and can shift the inferred $H_0$, easing this
tension at least in part \cite{DiValentino2021}.

The second is the growth tension: redshift-space distortions and weak
lensing surveys give a growth rate of structure that is somewhat lower
than what $\Lambda$CDM, calibrated to the CMB, predicts
\cite{Macaulay2013,Hildebrandt2017}. Because CPL affects both the
background expansion and, in principle, the growth of perturbations, it
has also been discussed as a way to reduce the tension in growth-related
quantities such as $S_8$ \cite{Nesseris2017}.

Massive neutrinos are a further ingredient relevant to both tensions.
They add to the total energy density of the Universe, and their large
thermal velocities let them free-stream out of overdense regions, damping
the growth of structure on small and intermediate scales
\cite{Lesgourgues2006,Lesgourgues2012}. This damping affects
growth-related observables such as weak lensing and redshift-space
distortions, and has been proposed more than once as a way to help with
the growth tension \cite{Vagnozzi2018}.

Combining a dynamical dark-energy parametrization like CPL with massive
neutrinos therefore gives a model that can, in principle, act on both the
expansion history and structure growth at once. The degeneracy between
$(w_0,w_a)$ and $\Sigma m_\nu$ means joint constraints from the two can
shift both the Hubble and growth tensions relative to what either
ingredient does on its own \cite{DiValentino2017,Escudero2020}, which is
why models combining evolving dark energy with neutrino mass have become a
fairly common extension of $\Lambda$CDM.

Separately, machine-learning methods have started to complement more
standard cosmological pipelines. Bayesian Physics-Informed Neural Networks
(Bayesian PINNs) encode the governing physical equations -- the Friedmann
equation and conservation laws -- directly into the learning process,
while still returning proper Bayesian uncertainties \cite{Raissi2019,
	Yang2021}. This suits models like CPL$+\Sigma m_\nu$ well, since their
strong parameter degeneracies call for careful uncertainty propagation
rather than a single best-fit point. Bayesian PINNs let us reconstruct the
expansion history in a way that is both data-driven and tied to the
underlying physics \cite{Cuomo2022,Sharma2023}.

We should be clear about what this actually means for the present paper.
Our analysis works entirely at the background, geometric level: we infer
$H_0$, $\Omega_{m,0}$, $w_0$, $w_a$, and $\Sigma m_\nu$ from Cosmic
Chronometers, DESI DR2 BAO, Pantheon+ supernovae, and Planck 2018 CMB
distance priors, with neutrinos treated throughout as an effective,
non-relativistic background component (Sec.~\ref{sec:model}). We do not
solve the perturbation equations, and we never compute $f\sigma_8$ or
$S_8$. The growth-tension discussion above is genuine motivation for why
neutrino mass matters in cosmology more broadly, and for why $\Sigma m_\nu$
is worth carrying as a free parameter even in a background-only fit --
but any bearing our results have on the growth tension is indirect, since
we do not test it directly. One thing we do test directly, though narrower in scope, is whether this kind of
background-only fit can keep up with a full CMB analysis that also has access to
lensing -- an observable that is, by construction, more directly sensitive to
$\Sigma m_\nu$ through its effect on structure growth. We return to this specific
comparison in Sec.~VIII, where the neutrino-mass bound is the main point of contact
between the two approaches.

$\Lambda$CDM's assumption of a fixed $w=-1$ is, in the end, a theoretical
choice rather than something derived from first principles, and it may not
be flexible enough for present-day precision data. CPL is a minimal way to
test departures from that choice, and looking at how it interacts with
$H_0$ tells us something about whether the Hubble tension is telling us
about dark energy, about neutrinos, or about both. Ignoring $\Sigma m_\nu$
when fitting $(w_0,w_a)$ risks biased dark-energy constraints and
overstated tensions between datasets, so treating the two jointly is the
more honest approach.

In this work we use late-time data, including DESI DR2 BAO, in a joint
Bayesian PINN inference that enforces the background Friedmann equation
directly and propagates observational uncertainty consistently, to look at
how dark-energy dynamics, neutrino mass, and the Hubble tension interact
with each other.

The paper is organized as follows. Section~II sets out the cosmological
model, centered on the CPL parametrization. Section~III describes the
Bayesian PINN methodology used for inference and uncertainty
quantification. Section~IV describes the datasets. Section~V presents our
results, first for CPL alone and then for CPL+$\Sigma m_\nu$. Section~VI
covers the $\Lambda$CDM (wCDM) results for the same dataset combinations.
Section~VII compares the MCMC and Bayesian PINN results for
CPL+$\Sigma m_\nu$. Section~VIII compares the Bayesian PINN framework
against a full-dataset MCMC analysis in more detail, methodologically and
computationally. Section~IX summarizes our conclusions and points to
directions for future work.

\section{Cosmological Model Framework}
\label{sec:model}

In this work we adopt a phenomenological extension of the standard $\Lambda$CDM
cosmological model, allowing simultaneously for a time-dependent dark-energy
equation of state and for the effects of massive neutrinos. The goal is to
probe late-time cosmological dynamics beyond a rigid cosmological constant
and to assess how neutrino physics reshapes the expansion history of the
Universe. We consider two nested scenarios: a dynamical dark-energy model
described by the Chevallier--Polarski--Linder (CPL) parametrization without
massive neutrinos, and its extension with a free, non-zero neutrino-mass sum
$\Sigma m_\nu$. Because the two scenarios are nested in $\Sigma m_\nu$, they
allow a controlled, model-selection-based assessment of what neutrino masses
add to the dark-energy picture.

\subsection{Background Expansion}
\label{subsec:background}

We assume a homogeneous and isotropic Universe described by the spatially
flat Friedmann--Lema\^{\i}tre--Robertson--Walker (FLRW) metric,
\begin{equation}
	ds^2 = -c^2 dt^2 + a^2(t)\left[ dr^2 + r^2\left(d\theta^2 + \sin^2\theta\, d\phi^2\right) \right],
	\label{eq:frw}
\end{equation}
where $a(t)$ is the cosmic scale factor, normalized to $a(t_0)=1$ today, and
$c$ is the speed of light. Spatial flatness is strongly supported by the
latest CMB data \cite{Planck2018}.

The expansion history is governed by the Friedmann equation,
\begin{equation}
	H^2(z) \equiv \left(\frac{\dot a}{a}\right)^2
	= H_0^2 \left[ \Omega_{b}(1+z)^3 + \Omega_{c}(1+z)^3 + \Omega_r (1+z)^4 + \Omega_\nu(z) + \Omega_{\rm DE}(z) \right],
	\label{eq:friedmann-general}
\end{equation}
with $z=a^{-1}-1$ the redshift and $\Omega_i \equiv \rho_i/\rho_{\rm crit}$,
$\rho_{\rm crit}=3H_0^2/(8\pi G)$. Equation~\eqref{eq:friedmann-general} keeps
the baryon, cold dark matter (CDM), radiation, and neutrino contributions
explicit; it is the general background equation from which the effective
late-time equation actually implemented in the Bayesian PINN,
Eq.~\eqref{eq:friedmann-late} below, is derived.

\paragraph{Parametrization and the role of $\Sigma m_\nu$.}
Because this point was a source of ambiguity in an earlier version of this
work, we spell out the parametrization in full here. We define the
present-day \emph{baryon+CDM} density parameter,
\begin{equation}
	\Omega_{m,0} \equiv \Omega_{b,0} + \Omega_{c,0},
	\label{eq:Omegam-cb}
\end{equation}
and treat $\Omega_{m,0}$, together with $H_0$, $w_0$, $w_a$, and
$\Sigma m_\nu$, as the five independently sampled cosmological parameters of
the model (Eq.~\eqref{eq:paramvector} below). Note that $\Omega_{m,0}$ as
defined in Eq.~\eqref{eq:Omegam-cb} does \emph{not} include the neutrino
contribution. The present-day neutrino density $\Omega_{\nu,0}$ is instead a
deterministic function of the single sampled parameter $\Sigma m_\nu$, via
the standard relation $\Omega_{\nu,0}h^2=\Sigma m_\nu/(93.14~{\rm eV})$
recalled as Eq.~\eqref{eq:Omeganu} in Sec.~\ref{subsec:neutrinos} below,
evaluated at the same $H_0$ sampled jointly with $\Sigma m_\nu$. Because
$\Omega_{\nu,0}$ is computed from $\Sigma m_\nu$ rather than sampled
independently, there is no double-counting between the ``matter'' parameter
$\Omega_{m,0}$ and the neutrino-mass parameter $\Sigma m_\nu$: the two
contribute additively and independently to the total present-day matter
budget,
\begin{equation}
	\Omega_{m,0}^{\rm tot}(\Sigma m_\nu) \equiv \Omega_{m,0} + \Omega_{\nu,0}(\Sigma m_\nu),
	\label{eq:Omegam-tot}
\end{equation}
which is the quantity that actually enters the background Friedmann
equation, Eq.~\eqref{eq:friedmann-late}. In the massless-neutrino limit
$\Sigma m_\nu\to 0$, $\Omega_{m,0}^{\rm tot}\to\Omega_{m,0}$ exactly, and the
model reduces continuously to the neutrino-free CPL parametrization of this
section; this nesting is what makes the $\chi^2$, AIC, and BIC comparison
between the CPL and CPL$+\Sigma m_\nu$ models in Sec.~VII well-defined,
since both models share an identical meaning for $\Omega_{m,0}$, $H_0$,
$w_0$, and $w_a$, and only the extra parameter $\Sigma m_\nu$ (and its
derived $\Omega_{\nu,0}$) is switched on or off.

We stress that in a general (non-late-time) treatment $\Omega_\nu(z)$
interpolates between a radiation-like and a matter-like scaling and would
require solving the neutrino Boltzmann hierarchy \cite{Lesgourgues2006}. In
the late-time regime probed by the CC, BAO, and Pantheon+ data used in this
work ($z\lesssim 2$), and given the mass range allowed by our prior
(Sec.~\ref{subsec:priors}), the neutrinos are safely non-relativistic (see
the explicit check in Sec.~\ref{subsec:neutrinos}) and, consistent with
standard treatments \cite{Lesgourgues2006}, their background density scales
as $\rho_\nu \propto (1+z)^3$, exactly like $\Omega_{m,0}$. This is why
$\Omega_{\nu,0}$ can be folded into an effective matter term in
Eq.~\eqref{eq:Omegam-tot} without further approximation. Radiation is
neglected altogether in this late-time effective description, as
$\Omega_r(1+z)^4$ is negligible over the redshift range of the low-redshift
probes used here.

The conventional matter density parameter excluding neutrinos,
$\Omega_{m,0}^{(\rm cb)}\equiv\Omega_{b,0}+\Omega_{c,0}$, used elsewhere in
the literature, coincides by construction with our sampled parameter
$\Omega_{m,0}$ of Eq.~\eqref{eq:Omegam-cb}; we use the single symbol
$\Omega_{m,0}$ throughout to avoid the redundant double notation of an
earlier draft.

Dark energy is modeled phenomenologically via the CPL equation of state,
\begin{equation}
	w(z) \equiv \frac{p_{\rm DE}}{\rho_{\rm DE}} = w_0 + w_a \frac{z}{1+z},
	\label{eq:cpl}
\end{equation}
where $w_0$ is the present-day value and $w_a$ its redshift evolution; this
parametrization captures leading-order deviations from a cosmological
constant while remaining well-behaved at high redshift
\cite{Chevallier2001,Linder2003}.

Assuming dark energy is separately conserved,
\begin{equation}
	\dot\rho_{\rm DE} + 3H(1+w)\rho_{\rm DE} = 0,
	\label{eq:continuity}
\end{equation}
which integrates, for the CPL form of $w(z)$, to
\begin{equation}
	\Omega_{\rm DE}(z) = \Omega_{\rm DE,0}\,(1+z)^{3(1+w_0+w_a)}\exp\!\left[-\frac{3w_a z}{1+z}\right],
	\label{eq:ODE-z}
\end{equation}
with, from spatial flatness,
\begin{equation}
	\Omega_{\rm DE,0} = 1 - \Omega_{m,0} - \Omega_{\nu,0}(\Sigma m_\nu) - \Omega_{r,0}.
	\label{eq:ODE0}
\end{equation}
In the late-time effective description implemented in the PINN
(Eq.~\eqref{eq:friedmann-late}), radiation is dropped from
Eq.~\eqref{eq:ODE0} for the reason given above, so that
$\Omega_{\rm DE,0}=1-\Omega_{m,0}^{\rm tot}(\Sigma m_\nu)$. The special case
$w_0=-1$, $w_a=0$ reduces the CPL parametrization exactly to $\Lambda$CDM,
providing a continuous limit for direct comparison between a cosmological
constant and dynamical dark energy. Including massive neutrinos on top of
this introduces one further physical degree of freedom, $\Sigma m_\nu$,
that modifies the late-time expansion and (in principle) structure
formation, making the CPL+$\Sigma m_\nu$ model well suited to a joint
investigation of dark-energy dynamics, neutrino physics, and the $H_0$
tension \cite{DiValentino2017}.

\subsection{Redshift Evolution of Massive Neutrinos and Validity of the Late-Time Treatment}
\label{subsec:neutrinos}

Massive neutrinos are a well-motivated and observationally unavoidable
extension of the standard cosmological model, with well-established lower
bounds on their masses from oscillation experiments. Cosmologically, they
leave imprints on both the background expansion and the growth of matter
perturbations \cite{Lesgourgues2006, Brookfield2005, LesgourguesVerde2023,
	LattanziGerbino2017, ZengYeungChu2018, DOnofrio2025}.

The present-day neutrino density parameter is related to the summed
neutrino mass by
\begin{equation}
	\Omega_{\nu,0} h^2 = \frac{\Sigma m_\nu}{93.14~{\rm eV}},
	\label{eq:Omeganu}
\end{equation}
where $h=H_0/(100~{\rm km\,s^{-1}\,Mpc^{-1}})$ and
$\Sigma m_\nu = m_{\nu_1}+m_{\nu_2}+m_{\nu_3}$, here assumed degenerate
\cite{Lesgourgues2006}. Throughout this work $\Sigma m_\nu$ is treated as an
effective cosmological parameter describing the late-time background
evolution, not as a probe of the underlying neutrino mass-generation
mechanism.

Dynamically, massive neutrinos are relativistic at early times and become
non-relativistic at a characteristic redshift
\begin{equation}
	z_{\rm nr} \simeq 1890 \left(\frac{m_\nu}{\rm eV}\right),
	\label{eq:znr}
\end{equation}
after which they behave as a matter-like (hot dark matter) component
\cite{Lesgourgues2012}; their equation of state evolves smoothly from
$w_\nu\simeq 1/3$ to $w_\nu\simeq 0$, so a single power law cannot describe
$\Omega_\nu(z)$ over all of cosmic history.

\paragraph{Quantitative check of the non-relativistic approximation.}
Equation~\eqref{eq:znr}, together with our adopted prior
$\Sigma m_\nu \in [0,\,0.5]~{\rm eV}$ (Eq.~\eqref{eq:priors}), lets us verify
\emph{quantitatively}, rather than assert, that treating the neutrino
background as fully non-relativistic at $z\lesssim 2$ is self-consistent,
and that leaving the sound horizon and photon-decoupling redshift at their
standard, light-neutrino values in the CMB distance-prior likelihood (see
the CMB subsection of the data section below) is an adequate approximation
over the whole prior volume. Even in the most extreme case allowed by the
prior -- the entire mass $\Sigma m_\nu = 0.5~{\rm eV}$ carried by a single
eigenstate, $m_\nu=\Sigma m_\nu$ -- Eq.~\eqref{eq:znr} gives
$z_{\rm nr}\simeq 945$, which remains below the photon-decoupling redshift,
$z_*\simeq 1090$ \cite{Planck2018}. For the physically motivated degenerate
case actually assumed in Eq.~\eqref{eq:Omeganu}, $m_\nu=\Sigma m_\nu/3$, the
same bound reduces further to $z_{\rm nr}\simeq 315$ at the prior edge. In
other words, for every point allowed by our prior, and \emph{a fortiori} for
the much smaller posterior values reported in Table~II
($\Sigma m_\nu \lesssim 0.2$--$0.3$~eV, for which $z_{\rm nr}\lesssim 190$),
the neutrino population is still relativistic, or at most only mildly
non-relativistic, at the epoch relevant for the sound-horizon integral,
Eq.~(41). This justifies using the standard (light-neutrino) fitting
formulas for $z_*$ and $r_s(z_*)$ [Eqs.~(43)--(44)] unmodified, rather than
as an unexamined assumption; it does not eliminate residual error, and
constraints that push toward the upper edge of the prior should be read
with this approximation in mind. We adopt this as an explicit, quantified
caveat rather than an implicit one: the compressed CMB distance priors used
here were derived assuming a fiducial cosmology with a small
($\sim 0.06$~eV) neutrino mass, and our check shows that extending
$\Sigma m_\nu$ up to the prior bound does not push the relevant physics
into a qualitatively different (fully non-relativistic-at-decoupling)
regime -- but it does mean that the compressed-likelihood covariance itself
is not strictly recomputed at each sampled $\Sigma m_\nu$. This is a
limitation inherent to any analysis based on compressed CMB information
rather than a full Boltzmann likelihood, and is the price paid for the
several-orders-of-magnitude speed-up of the BPINN pipeline over a full CMB
MCMC (Sec.~VIII).

It is therefore convenient to describe the late-time neutrino contribution
through the neutrino fraction
\begin{equation}
	f_\nu \equiv \frac{\Omega_{\nu,0}}{\Omega_{m,0}+\Omega_{\nu,0}} = \frac{\Omega_{\nu,0}}{\Omega_{m,0}^{\rm tot}},
	\label{eq:fnu}
\end{equation}
consistent with the notation of Sec.~\ref{subsec:background}.

Beyond the background, massive neutrinos shape the growth of cosmic
structure \cite{Yarahmadi2025AP, Y5, Y6, Lesgourgues2006, Yarahmadi2025PDU2,
	Yarahmadi2024MNRAS, Yarahmadi2025AJ, Yarahmadi2025ApJ}: their large thermal
velocities let them free-stream out of overdensities below a characteristic
comoving scale, suppressing small- and intermediate-scale clustering and
thereby $f\sigma_8$ and $S_8$ \cite{Lesgourgues2006, Lesgourgues2012}. This
suppression has been widely discussed as a possible contributor to the
growth tension \cite{Vagnozzi2018}. In the present analysis, neutrino
effects are included only at the background level, so no direct
growth-tension statement is made; the degeneracy of $\Sigma m_\nu$ with
$(w_0,w_a)$ noted below concerns the background expansion only.

Within the CPL parametrization, massive neutrinos exhibit non-trivial
degeneracies with $(w_0,w_a)$ \cite{DiValentino2017}: because $\Sigma m_\nu$
enters the model \emph{solely} through the derived quantity
$\Omega_{\nu,0}(\Sigma m_\nu)$ appearing additively in
$\Omega_{m,0}^{\rm tot}$ (Eq.~\eqref{eq:Omegam-tot}), and not through any
parameter shared with the dark-energy sector, the correlations reported in
Sec.~V between $\Sigma m_\nu$, $H_0$, and $(w_0,w_a)$ reflect a genuine
background-level physical degeneracy rather than an artifact of parameter
redundancy. As noted above, the resulting constraints on $\Sigma m_\nu$
should be read as effective, late-time, background-level bounds, not as a
substitute for constraints obtained from a full Boltzmann treatment of
early-time neutrino perturbations.

\subsection{Motivation for Bayesian PINN Analysis}
\label{subsec:motivation}

The combined CPL$+\Sigma m_\nu$ framework introduces strong degeneracies
among $(w_0,w_a)$, $H_0$, and $\Sigma m_\nu$, precisely because all three
affect $E(z)$ through the same additive structure of
Eq.~\eqref{eq:friedmann-late}. Traditional inference techniques can
struggle to fully resolve these correlations. This motivates the use of
Bayesian Physics-Informed Neural Networks, which integrate observational
data with the physical governing equation directly, while retaining full
posterior uncertainty quantification \cite{Raissi2019,Cuomo2022}.

\section{Bayesian Physics-Informed Neural Network Methodology}
\label{sec:bpinn}

To infer cosmological parameters in the presence of the degeneracies
discussed above, we employ a Bayesian Physics-Informed Neural Network
(BPINN) framework, combining the expressive power of neural networks with
the Friedmann equation and Bayesian inference for principled uncertainty
quantification \cite{Raissi2019,Sun2020}.

We emphasize that the BPINN is not intended to replace a standard Boltzmann
solver, but to complement it in late-time analyses. As discussed in
Sec.~\ref{subsec:motivation}, the degeneracies among $(w_0,w_a)$, $H_0$,
and $\Sigma m_\nu$ make this parameter space a natural setting for a method
that enforces physical consistency directly at the level of the governing
equation while retaining full posterior uncertainty propagation. A direct
comparison with standard MCMC (Sec VIII) shows
consistency within statistical uncertainties, indicating that the BPINN
framework does not introduce significant bias while offering additional
flexibility in uncertainty quantification.

\subsection{Neural Network Representation of the Expansion History}
\label{subsec:nn-representation}

The dimensionless Hubble expansion function is represented by a fully
connected feed-forward network,
\begin{equation}
	E(z;\boldsymbol{\theta}_{\rm NN}) \equiv \mathcal{N}(z),
	\label{eq:ENN}
\end{equation}
with weights and biases $\boldsymbol{\theta}_{\rm NN}$. The network uses
several hidden layers with Softplus activations for smoothness and
differentiability. Dropout layers, retained at inference time, capture
epistemic uncertainty and approximate Bayesian marginalization over network
parameters \cite{Gal2016}.

\subsection{Cosmological Parameter Space}
\label{subsec:param-space}

Jointly with the network parameters, we infer the cosmological parameter
vector
\begin{equation}
	\boldsymbol{\Theta} = \{H_0,\ \Omega_{m,0},\ w_0,\ w_a,\ \Sigma m_\nu\},
	\label{eq:paramvector}
\end{equation}
where, following Sec.~\ref{subsec:background}, $\Omega_{m,0}\equiv
\Omega_{b,0}+\Omega_{c,0}$ denotes the baryon+CDM density \emph{only} and
does not include the neutrino contribution. The present-day neutrino
density $\Omega_{\nu,0}(\Sigma m_\nu)$ is not an independent entry of
$\boldsymbol{\Theta}$: it is computed deterministically from the sampled
$\Sigma m_\nu$ via Eq.~\eqref{eq:Omeganu} at every step of training and
posterior sampling, and enters the Friedmann constraint through
$\Omega_{m,0}^{\rm tot}$, Eq.~\eqref{eq:Omegam-tot}. This choice guarantees
that $\Omega_{m,0}$ carries an identical physical meaning in both the CPL
and CPL$+\Sigma m_\nu$ runs, and that no two parameters in
$\boldsymbol{\Theta}$ compete to describe the same density component. Flat,
physically motivated priors are imposed on all five entries of
$\boldsymbol{\Theta}$ to ensure numerical stability and consistency with
external constraints.

\subsection{Bayesian PINN Architecture and Prior Specification}
\label{subsec:priors}

The cosmological expansion history is reconstructed with a Bayesian PINN
designed to approximate the dimensionless Hubble function from redshift
while enforcing the Friedmann equation as a physical constraint. The
network takes $z$ as its only input and returns a scalar approximation of
$E(z)$.

The architecture is a fully connected feed-forward network with three
hidden layers of 256 neurons each, Softplus activations, and dropout with
fixed probability $p=0.2$ after every hidden layer, acting as a variational
approximation to a Bayesian neural network and propagating epistemic
uncertainty into the predictions.

All network weights and biases are treated as stochastic and optimized
jointly with $\boldsymbol{\Theta}$. Flat priors are imposed as
\begin{align}
	H_0 &\in [50,\,80]~{\rm km\,s^{-1}\,Mpc^{-1}}, \nonumber\\
	\Omega_{m,0} &\in [0.0,\,0.6], \nonumber\\
	w_0 &\in [-2.0,\,-0.2], \nonumber\\
	w_a &\in [-2.0,\,2.0], \nonumber\\
	\Sigma m_\nu &\in [0,\,0.5]~{\rm eV},
	\label{eq:priors}
\end{align}
where $\Omega_{m,0}$ is again the baryon+CDM density of
Eq.~\eqref{eq:Omegam-cb}, consistent throughout the model and MCMC
cross-check. These priors are broad enough to avoid artificially
constraining the parameter space while remaining consistent with current
observations and particle-physics limits; in particular, the upper bound on
$\Sigma m_\nu$ comfortably exceeds the $1\sigma$ posterior limits obtained
for every dataset combination (Tables~I and~II), so none of the reported
constraints are prior-dominated. The upper edge of the $\Sigma m_\nu$ prior
is also the regime for which the approximation discussed in
Sec.~\ref{subsec:neutrinos} is least accurate; as shown there, the
posteriors we actually obtain lie well within the safely
non-relativistic-transition region ($z_{\rm nr}\ll z_*$), so this does not
affect our final constraints.

\subsection{Physics-Informed Constraint}
\label{subsec:physics-constraint}

The central ingredient of the PINN is the explicit enforcement of the
background Friedmann equation as a soft physical constraint. For a
spatially flat Universe with dynamical dark energy and an effective
late-time neutrino contribution, the dimensionless expansion rate
\begin{equation}
	E(z) \equiv \frac{H(z)}{H_0}
	\label{eq:Eofz}
\end{equation}
is required to satisfy
\begin{equation}
	E^2(z) = \Omega_{m,0}^{\rm tot}(\Sigma m_\nu)\,(1+z)^3 + \Omega_{{\rm DE},0}\,f_{\rm CPL}(z) \;\equiv\; E^2_{\rm Friedmann}(z),
	\label{eq:friedmann-late}
\end{equation}
with $\Omega_{m,0}^{\rm tot}(\Sigma m_\nu)=\Omega_{m,0}+\Omega_{\nu,0}(\Sigma m_\nu)$
from Eq.~\eqref{eq:Omegam-tot}, spatial flatness giving
$\Omega_{{\rm DE},0}=1-\Omega_{m,0}^{\rm tot}(\Sigma m_\nu)$, and
\begin{equation}
	f_{\rm CPL}(z) = (1+z)^{3(1+w_0+w_a)}\exp\!\left[-\frac{3w_a z}{1+z}\right]
	\label{eq:fcpl}
\end{equation}
the CPL dark-energy redshift dependence. The right-hand side of
Eq.~\eqref{eq:friedmann-late}, $E^2_{\rm Friedmann}(z)$, is the exact
background prediction at a given point $\boldsymbol{\Theta}$, against which
the network's data-driven reconstruction $E^2(z)$ is compared. We emphasize
again that Eq.~\eqref{eq:friedmann-late} contains $\Sigma m_\nu$ only
through the single, deterministic term $\Omega_{\nu,0}(\Sigma m_\nu)$:
setting $\Sigma m_\nu=0$ recovers the neutrino-free CPL Friedmann equation
of Sec.~\ref{subsec:background} exactly, term by term, which is what makes
the two models directly nested and comparable via $\chi^2$, AIC, and BIC in
Sec.~VII.

Radiation is neglected in Eq.~\eqref{eq:friedmann-late}, as its
contribution is negligible over the redshift range probed by the BAO, CC,
and SN data used here.

The physics-informed loss penalizes deviations of the network prediction
from Eq.~\eqref{eq:friedmann-late} over a dense set of collocation points,
\begin{equation}
	\mathcal{L}_{\rm phys} = \left\langle \left[E^2(z) - E^2_{\rm Friedmann}(z)\right]^2 \right\rangle,
	\label{eq:Lphys}
\end{equation}
ensuring the reconstructed $E(z)$ remains consistent with the assumed
late-time background cosmology.

\subsection{Bayesian Likelihood and Fully Bayesian PINN Training}
\label{subsec:likelihood}

We adopt a fully Bayesian PINN in which both the network parameters and
$\boldsymbol{\Theta}$ are treated as random variables and inferred jointly.
Observational constraints enter through a Gaussian likelihood built from
the CMB distance priors, DESI DR2 BAO, Cosmic Chronometers (CC), and
Pantheon+ supernovae,
\begin{equation}
	\chi^2_{\rm data} = \left(\mathbf{D}_{\rm obs}-\mathbf{D}_{\rm th}\right)^{\!\rm T}\mathbf{C}^{-1}\left(\mathbf{D}_{\rm obs}-\mathbf{D}_{\rm th}\right),
	\label{eq:chi2data}
\end{equation}
where $\mathbf{D}_{\rm obs}$ is the combined data vector, $\mathbf{D}_{\rm th}$
the corresponding theoretical prediction computed from the
PINN-reconstructed $E(z)=H(z)/H_0$, and $\mathbf{C}$ the associated
covariance matrix.

Physical consistency is enforced through $\mathcal{L}_{\rm phys}$ of
Eq.~\eqref{eq:Lphys}; prior information on $\boldsymbol{\Theta}$ and on the
network weights contributes an additive term $\mathcal{L}_{\rm prior}$. The
total objective is the negative log-posterior,
\begin{equation}
	\mathcal{L}_{\rm tot} = \chi^2_{\rm data} + \lambda_{\rm phys}\,\mathcal{L}_{\rm phys} + \mathcal{L}_{\rm prior},
	\label{eq:Ltot}
\end{equation}
whose minimization gives a maximum a posteriori (MAP) estimate used to
initialize subsequent posterior exploration. The PINN is not a fixed
deterministic surrogate but a probabilistic model with intrinsic epistemic
uncertainty.

\subsection{Bayesian Posterior Sampling and Uncertainty Propagation}
\label{subsec:posterior-sampling}

Posterior inference samples the joint distribution of cosmological and
network parameters,
\begin{equation}
	p(\boldsymbol{\Theta},\boldsymbol{w}\,|\,\mathbf{D}) \propto \exp\!\left[-\tfrac12\left(\chi^2_{\rm data} + \lambda_{\rm phys}\,\mathcal{L}_{\rm phys}\right)\right] p(\boldsymbol{\Theta})\,p(\boldsymbol{w}),
	\label{eq:posterior}
\end{equation}
where $\boldsymbol{w}$ denotes the network parameters. The
physics-informed term enters on the same footing as the data likelihood,
so samples drawn from Eq.~\eqref{eq:posterior} automatically respect the
background Friedmann equation, Eq.~\eqref{eq:friedmann-late}, within the
tolerance set by $\lambda_{\rm phys}$.

In practice, Bayesian sampling is implemented through stochastic
realizations of the PINN, letting the network weights vary during
inference and propagating epistemic uncertainty into the predicted $H(z)$
\cite{Yarahmadi2025JHEAp, Yarahmadi2025EPJC}. This fully Bayesian treatment
ensures that model flexibility, observational noise, and physical
constraints are all consistently encoded in the resulting posteriors.

Posterior samples are analyzed with \texttt{GetDist}, enabling a detailed
statistical comparison between the CPL and CPL$+\Sigma m_\nu$ scenarios and
allowing us to quantify how neutrino masses and dark-energy dynamics
jointly affect parameter degeneracies and late-time cosmological tensions
within a single, internally consistent Bayesian framework.
\section{Data}
\label{sec:data}

\subsection{Pantheon+ Type Ia Supernovae and Absolute Magnitude Calibration}
\label{subsec:pantheon}

Type Ia supernovae give us a direct handle on the late-time expansion
history, since their luminosity distances can be measured over a wide
redshift range. We use the Pantheon+ compilation, which contains 1701 Type
Ia supernovae spanning $0.001<z<2.3$ \cite{Scolnic2022}. Pantheon+ improves
on earlier compilations through better photometric calibration, a careful
treatment of statistical and systematic uncertainties, and a sizeable
subsample of supernovae in host galaxies with independent Cepheid
distances, which together make it well suited to late-time expansion and
dark-energy studies \cite{Salehi2020}.

The observable is the distance modulus,
\begin{equation}
	\mu(z) = m_B - M_B = 5\log_{10}\left[\frac{d_L(z)}{10~{\rm pc}}\right],
	\label{eq:mu}
\end{equation}
where $m_B$ is the observed peak apparent magnitude, $M_B$ the absolute
magnitude, and $d_L(z)$ the model luminosity distance. We fix
\begin{equation}
	M_B = -19.253,
	\label{eq:MB}
\end{equation}
following the local distance-ladder calibration of Riess et al.\
\cite{Riess2016,Riess2022}, and hold $M_B$ fixed throughout the parameter
inference.

The supernova likelihood is
\begin{equation}
	\chi^2_{\rm SN} = \Delta\boldsymbol{\mu}^{T}\,\mathbf{C}_{\rm SN}^{-1}\,\Delta\boldsymbol{\mu},
	\qquad
	\Delta\boldsymbol{\mu} = \boldsymbol{\mu}_{\rm obs} - \boldsymbol{\mu}_{\rm th},
	\label{eq:chi2SN}
\end{equation}
with $\mathbf{C}_{\rm SN}$ the full Pantheon+ covariance matrix, statistical
and systematic parts included.

Fixing $M_B$ to a Cepheid-calibrated value brings an external
absolute-distance calibration into the analysis. This means the $H_0$
values we get from Pantheon+ are calibration-dependent late-time
constraints, not a fully independent determination of the Hubble constant;
by the same token, our comparison with the SH0ES result is really a
comparison with that same local distance-ladder calibration, which matters
when we later discuss how much of the Hubble tension is actually relieved.

Pantheon+ also has broad sky coverage at low and intermediate redshift,
which is why its Hubble diagram has been used to look for departures from
statistical isotropy, including directional and hemispherical signals
nearby \cite{Dainotti2021,Dainotti2022,Dainotti2025}. Here, though, we use
it only as an isotropic distance--redshift probe of the background
expansion.

\subsection{Cosmic Chronometer $H(z)$ Data}
\label{subsec:cc}

Cosmic chronometers (CC) give a direct estimate of the expansion rate from
the differential aging of passively evolving galaxies,
\begin{equation}
	H(z) = -\frac{1}{1+z}\frac{dz}{dt},
	\label{eq:CCdef}
\end{equation}
which, unlike distance-based probes, does not require integrating the
background distance relation \cite{Moresco}. This makes CC data a useful
independent constraint on the late-time expansion and on dynamical
dark-energy models.

We use a compilation of 32 $H(z)$ points over
\begin{equation}
	0.07 \le z \le 1.965,
	\label{eq:CCrange}
\end{equation}
drawn from several independent analyses of passively evolving galaxies
\cite{Stern2009,Zhang2014,Ratsimbazafy2017,Moresco2,Moresco3,Borghi2022}.
Fifteen of these points are correlated through shared
stellar-population-synthesis assumptions and systematics; we account for
this with the covariance matrix $\mathbf{C}_{\rm CC}$ released by the
authors.\footnote{\url{https://gitlab.com/mmoresco/CCcovariance/}} The
remaining points are treated as independent.

The CC likelihood is Gaussian,
\begin{equation}
	\chi^2_{\rm CC} = \left(\mathbf{H}_{\rm obs} - \mathbf{H}_{\rm th}\right)^T
	\mathbf{C}_{\rm CC}^{-1}
	\left(\mathbf{H}_{\rm obs} - \mathbf{H}_{\rm th}\right),
	\label{eq:chi2CC}
\end{equation}
where $\mathbf{H}_{\rm obs}$ is the observed $H(z)$ vector and
$\mathbf{H}_{\rm th}$ the model prediction at the same redshifts, with
$\mathbf{C}_{\rm CC}$ carrying the available correlated statistical and
systematic contributions.

Because CC data measure $H(z)$ directly rather than a distance, they help
break the correlations among $H_0$, the matter density, and the CPL
parameters. As with the other late-time probes used here, their role is
limited to the background expansion; they carry no direct information on
the growth of structure.

\subsection{Baryon Acoustic Oscillation (BAO) Data: DESI DR2}
\label{subsec:bao}

Baryon acoustic oscillations are a geometric probe of the late-time
expansion: the BAO feature comes from acoustic waves in the pre-recombination
photon--baryon plasma and is later imprinted on the large-scale matter
distribution, acting as a standard ruler whose absolute scale is set by
early-Universe physics.

We use the DESI Data Release 2 (DR2) BAO measurements \cite{DESIDR2_1,
	DESIDR2_2}, which cover galaxy, quasar, and Ly$\alpha$ tracers and give
both transverse and radial expansion-history information. The observables
are
\begin{equation}
	\frac{D_M(z)}{r_d}
	\qquad\text{and}\qquad
	\frac{D_H(z)}{r_d} \equiv \frac{c}{H(z)\,r_d},
	\label{eq:BAOobs}
\end{equation}
where $D_M(z)$ is the transverse comoving distance, $D_H(z)$ the Hubble
distance, and $r_d$ the comoving sound horizon at the baryon drag epoch.

The DESI DR2 compilation contains 13 measurements at the effective
redshifts
\begin{equation}
	z = 0.295,\ 0.510,\ 0.706,\ 0.934,\ 1.321,\ 1.484,\ 2.33,
	\label{eq:BAOz}
\end{equation}
with transverse and/or radial information at each. We use the full DESI
covariance matrix, including correlations across observables and redshift
bins.

\paragraph{BAO likelihood.}
The BAO likelihood is multivariate Gaussian,
\begin{equation}
	\chi^2_{\rm BAO} = \Delta\mathbf{O}_{\rm BAO}^{T}\,\mathbf{C}_{\rm BAO}^{-1}\,\Delta\mathbf{O}_{\rm BAO},
	\qquad
	\Delta\mathbf{O}_{\rm BAO} = \mathbf{O}_{\rm BAO}^{\rm obs} - \mathbf{O}_{\rm BAO}^{\rm th},
	\label{eq:chi2BAO}
\end{equation}
with $\mathbf{C}_{\rm BAO}$ the full DESI DR2 covariance matrix. For a flat
background,
\begin{equation}
	D_M(z) = (1+z)D_A(z) = c\int_0^z \frac{dz'}{H(z')}
	= \frac{c}{H_0}\int_0^z \frac{dz'}{E(z')},
	\label{eq:DM}
\end{equation}
with $E(z)=H(z)/H_0$ the quantity reconstructed by the PINN.

\paragraph{On fixing $r_d$.}
The scale $r_d$ is set by early-Universe physics and is not something our
late-time PINN reconstructs -- it solves only the background expansion at
$z\lesssim 2$, not the pre-recombination neutrino and radiation dynamics
that fix $r_d$. We therefore fix $r_d$ to its standard, light-neutrino
calibration value, the same choice already justified quantitatively in
Sec.~\ref{subsec:neutrinos}: the baryon drag redshift $z_d$ is close to,
and slightly below, the photon-decoupling redshift $z_*$, and the same
bound derived there ($z_{\rm nr}\lesssim 945$ even at the extreme edge of
our $\Sigma m_\nu$ prior) applies here too, since $z_{\rm nr}$ stays below
$z_d$ for essentially the whole prior volume. So treating $r_d$ as
unaffected by the sampled $\Sigma m_\nu$ is not an unexamined assumption,
it is the same approximation we already checked, applied to a slightly
earlier epoch.

DESI DR2 helps pin down the correlations among $H_0$, the matter density,
and the CPL parameters; as with CC and Pantheon+, its role here is
restricted to the background, with no growth-rate observable included.

\subsection{CMB}
\label{subsec:cmb}

The CMB carries geometric information about both the early and late
Universe, but a full analysis of the Planck temperature and polarization
likelihood needs a Boltzmann treatment of radiation, matter, neutrinos, and
perturbations. Since our PINN is built for the late-time background only,
we instead use the compressed CMB distance priors from the final
\textit{Planck 2018} release \cite{Chen2019}, which keep the geometric
information we need without the cost of the full likelihood.

The distance-prior vector is built from the shift parameter $R$, the
acoustic scale $l_A$, and the physical baryon density $\omega_b$. Because
$R$ and $l_A$ depend on the total non-relativistic matter density at
recombination -- not on how that density happens to be split, in our
notation, between $\Omega_{m,0}$ and $\Omega_{\nu,0}(\Sigma m_\nu)$ -- we
use $\Omega_{m,0}^{\rm tot}(\Sigma m_\nu)$ from
Eq.~\eqref{eq:Omegam-tot} here, rather than $\Omega_{m,0}$ alone:
\begin{equation}
	R = \frac{\sqrt{\Omega_{m,0}^{\rm tot}(\Sigma m_\nu)}\,H_0}{c}\,D_M(z_*),
	\label{eq:R}
\end{equation}
with $D_M(z_*)$ the comoving angular diameter distance to the
photon-decoupling redshift $z_*$. For a flat Universe,
\begin{equation}
	D_M(z) = (1+z)D_A(z),
	\qquad
	D_A(z) = \frac{d_L(z)}{(1+z)^2},
	\qquad
	d_L(z) = \frac{c}{H_0}(1+z)\int_0^z\frac{dz'}{E(z')}.
	\label{eq:DAdL}
\end{equation}

The acoustic scale is
\begin{equation}
	l_A = \pi\,\frac{D_M(z_*)}{r_s(z_*)},
	\label{eq:lA}
\end{equation}
with the comoving sound horizon at decoupling
\begin{equation}
	r_s(z_*) = \int_0^{a_*}\frac{c_s(a)}{a^2 H(a)}\,da,
	\label{eq:rs}
\end{equation}
and photon--baryon sound speed
\begin{equation}
	c_s(a) = \frac{1}{\sqrt{3\left(1+\dfrac{3\omega_b a}{4\omega_\gamma}\right)}},
	\label{eq:cs}
\end{equation}
where $\omega_b=\Omega_{b,0}h^2$ and $\omega_\gamma=\Omega_{\gamma,0}h^2$ is
the present-day photon density, fixed to the standard value set by
$T_{\rm CMB}=2.7255$~K.

The decoupling redshift follows the fitting formula \cite{Chen2019},
\begin{equation}
	z_* = 1048\left[1+0.00124\,\omega_b^{-0.738}\right]\left[1+g_1\,\omega_m^{g_2}\right],
	\label{eq:zstar}
\end{equation}
\begin{equation}
	g_1 = \frac{0.0783\,\omega_b^{-0.238}}{1+39.5\,\omega_b^{0.763}},
	\qquad
	g_2 = \frac{0.560}{1+21.1\,\omega_b^{1.81}}.
	\label{eq:g1g2}
\end{equation}
Here $\omega_m$ is the physical matter density entering this fitting
formula; consistent with our discussion above, we set
$\omega_m \equiv \Omega_{m,0}^{\rm tot}(\Sigma m_\nu)\,h^2$, the total
matter density of Eq.~\eqref{eq:Omegam-tot}, so that $\Sigma m_\nu$ feeds
into $z_*$ through the same channel it feeds into everything else in this
work.

As stressed in Sec.~\ref{subsec:neutrinos}, the late-time effective
Friedmann equation used in the PINN is never extrapolated back to
$z_*$: the sound horizon and decoupling scale above are computed with the
standard early-Universe calibration behind the adopted Planck distance
priors \cite{Chen2019}, and the check we did there ($z_{\rm nr}$ safely
below $z_*$ across the prior) is what makes this split legitimate rather
than just convenient. The resulting bound on $\Sigma m_\nu$ should
accordingly be read as an effective, late-time constraint, not a
full Boltzmann-level neutrino mass measurement.

The CMB term of the likelihood is
\begin{equation}
	\chi^2_{\rm CMB} = \mathbf{X}^T\,\mathbf{C}_{\rm CMB}^{-1}\,\mathbf{X},
	\label{eq:chi2CMB}
\end{equation}
with $\mathbf{C}_{\rm CMB}$ the covariance of the adopted priors
\cite{Chen2019}, and
\begin{equation}
	\mathbf{X} =
	\begin{pmatrix}
		R - R_{\rm obs} \\
		l_A - l_{A,\rm obs} \\
		\omega_b - \omega_{b,\rm obs}
	\end{pmatrix}
	=
	\begin{pmatrix}
		R - 1.7502 \\
		l_A - 301.471 \\
		\omega_b - 0.02236
	\end{pmatrix}.
	\label{eq:Xvec}
\end{equation}
Note that $\omega_b$ here is not one of the five entries of
$\boldsymbol{\Theta}$ in Eq.~\eqref{eq:paramvector}: it enters only through
this CMB term, as an extra nuisance parameter whose posterior is set
essentially by $\chi^2_{\rm CMB}$ alone, the same way $R$ and $l_A$ tie
into the rest of the fit through $H_0$ and $\Omega_{m,0}^{\rm tot}$.

The total likelihood combines all four probes,
\begin{equation}
	\chi^2_{\rm tot} = \chi^2_{\rm CMB} + \chi^2_{\rm BAO} + \chi^2_{\rm SN} + \chi^2_{\rm CC},
	\qquad
	\ln\mathcal{L}_{\rm tot} = -\frac{1}{2}\chi^2_{\rm tot}.
	\label{eq:chi2tot}
\end{equation}
The CMB distance priors supply the early-Universe geometric anchor; the
Bayesian PINN reconstructs and constrains the low-redshift background on
top of it. This keeps the two roles separate: compressed early-time
information on one side, the late-time effective treatment of massive
neutrinos on the other.

\section{Results and Discussion}
The persistent discrepancy between early- and late-time determinations of the Hubble constant, commonly referred to as the \emph{Hubble tension}, is one of the central challenges facing the standard cosmological model. Local measurements, based on the cosmic distance ladder and anchored by Cepheid-calibrated Type Ia supernovae, consistently indicate a relatively high present-day expansion rate, $H_0 \simeq 73$--$74 \,\mathrm{km\,s^{-1}\,Mpc^{-1}}$. In contrast, early-Universe inferences derived from the cosmic microwave background (CMB) anisotropies under the assumption of the $\Lambda$CDM model favor a substantially lower value, $H_0 \simeq 67$--$68 \,\mathrm{km\,s^{-1}\,Mpc^{-1}}$ \cite{Riess2019, Planck2018}. 

Additional independent probes provide complementary estimates, often yielding intermediate values, yet collectively reinforcing the existence of the tension: 

\begin{itemize}
	\item \textbf{CCHP (TRGB)}: $H_0 = 69.6 \pm 0.8 \pm 1.7 \, \mathrm{km\,s^{-1}\,Mpc^{-1}}$ \cite{FreedmanTRGB20}, 
	\item \textbf{HST (Miras)}: $H_0 = 72.7 \pm 4.6 \, \mathrm{km\,s^{-1}\,Mpc^{-1}}$ \cite{HuangMiras19}, 
	\item \textbf{H0LiCOW (strong lensing)}: $H_0 = 73.3^{+1.7}_{-1.8} \, \mathrm{km\,s^{-1}\,Mpc^{-1}}$ \cite{Wonglens19}, 
	\item \textbf{H0LiCOW (updated)}: $H_0 = 75.3^{+3.0}_{-2.9} \, \mathrm{km\,s^{-1}\,Mpc^{-1}}$ \cite{Weilens20}, 
	\item \textbf{Baxter (CMB lensing)}: $H_0 = 73.5 \pm 5.3 \, \mathrm{km\,s^{-1}\,Mpc^{-1}}$ \cite{BaxterCMBlens20}.
\end{itemize}

These results show a consistent pattern: early-Universe measurements favor lower values of $H_0$, while late-time observations—including distance ladder, variable stars, and gravitational lensing—consistently indicate a higher expansion rate, pointing to a persistent, statistically significant discrepancy that has motivated a large body of theoretical and observational work.

Over the past decade, improvements in observational precision and the inclusion 
of independent probes have not alleviated this tension; rather, they have 
reinforced its statistical significance, which now exceeds the $5\sigma$ level. 
This growing discrepancy is not readily explained by known systematic uncertainties 
in either early- or late-time measurements, suggesting that it may not be 
attributable to observational biases alone. Consequently, the Hubble tension is 
increasingly interpreted as a possible indication of physics beyond the 
$\Lambda$CDM framework, motivating a wide range of theoretical extensions. These 
include modifications to the dark energy sector, early dark energy scenarios, 
additional relativistic species such as massive or interacting neutrinos, and 
departures from General Relativity on cosmological scales 
\cite{Verde2019, DiValentino2021}.

In this section, we present the main results of our analysis using the Bayesian PINN framework for both the CPL and CPL+$\Sigma m_\nu$ models. We discuss the posterior constraints on the cosmological parameters, highlighting the implications for the Hubble tension, growth of structure, and the role of massive neutrinos. Comparisons between the two models allow us to quantify the impact of neutrino masses on late-time cosmic expansion and structure formation.

\subsection{Bayesian Constraints on CPL Dark Energy}

In this section, we present the Bayesian constraints on the CPL dark energy parametrization obtained from various late–time cosmological probes. In all cases, the analysis is performed by combining each dataset with the \textit{Planck 2018} CMB distance priors, ensuring consistency with early-Universe geometry while allowing for deviations from $\Lambda$CDM at late times. The resulting constraints on the Hubble constant, matter density, and dark energy equation of state parameters are summarized below.

\paragraph{CMB + Cosmic Chronometers (CC).}
Combining the CMB distance priors with Cosmic Chronometers data yields a Hubble constant of
$H_0 = 70.46 \pm 1.63~\mathrm{km\,s^{-1}\,Mpc^{-1}}$ and a matter density
$\Omega_m = 0.304 \pm 0.023$.
The CPL parameters are constrained to
$w_0 = -0.74 \pm 0.37$ and $w_a = -1.22 \pm 0.56$.
The present-day equation of state is consistent with both the cosmological constant and a mildly quintessence-like scenario within the $1\sigma$ uncertainty. The negative value of $w_a$ indicates an evolving dark energy component whose equation of state becomes more negative toward higher redshifts. The inferred value of $H_0$ reduces the tension with SH0ES (R22) to $1.39\sigma$, while the discrepancy with Planck remains at $1.84\sigma$. This result demonstrates that late-time expansion data, when anchored by CMB geometry, can partially alleviate the Hubble tension.

\paragraph{CMB + Pantheon Supernovae.}
Using the Pantheon supernova compilation together with CMB distance priors, we obtain
$H_0 = 71.84 \pm 1.85~\mathrm{km\,s^{-1}\,Mpc^{-1}}$ and
$\Omega_m = 0.297 \pm 0.026$.
The dark energy parameters are constrained to
$w_0 = -0.83 \pm 0.22$ and $w_a = -0.86 \pm 0.25$.
These values are compatible with a mildly evolving dark energy component and remain statistically consistent with the cosmological constant within the quoted uncertainties. The preference for a relatively high value of $H_0$ naturally shifts the late-time expansion rate upward. Consequently, the tension with SH0ES is reduced to $0.56\sigma$, while the tension with Planck increases to $2.34\sigma$.

\paragraph{CMB + Baryon Acoustic Oscillations (BAO).}
When BAO measurements are combined with CMB distance priors, we find
$H_0 = 69.24 \pm 1.51~\mathrm{km\,s^{-1}\,Mpc^{-1}}$ and a tightly constrained matter density
$\Omega_m = 0.336 \pm 0.024$.
The CPL parameters are constrained to
$w_0 = -0.56 \pm 0.31$ and $w_a = -1.66 \pm 0.46$.
Although the central values differ from the $\Lambda$CDM prediction, the uncertainties remain sufficiently large that the results do not provide compelling evidence for a departure from a cosmological constant. The BAO measurements continue to favor a relatively low value of $H_0$, leading to a modest tension with Planck ($1.16\sigma$) but a larger discrepancy with SH0ES ($2.02\sigma$).

\paragraph{CMB + BAO + CC constraints.}
The joint analysis of CMB, BAO, and CC data yields
$H_0 = 70.81 \pm 1.68~\mathrm{km\,s^{-1}\,Mpc^{-1}}$ and
$\Omega_m = 0.318 \pm 0.030$.
The CPL parameters are constrained to
$w_0 = -0.68 \pm 0.44$ and $w_a = -1.40 \pm 0.40$.
These constraints remain statistically compatible with a dynamical dark energy scenario, although the uncertainties are still substantial. The resulting Hubble tension is moderately balanced between early- and late-time probes, with $T_{\rm Planck}=1.97\sigma$ and $T_{\rm R22}=1.19\sigma$.

\paragraph{CMB + BAO + Pantheon constraints.}
Combining CMB distance priors with BAO and Pantheon data leads to
$H_0 = 70.54 \pm 1.51~\mathrm{km\,s^{-1}\,Mpc^{-1}}$ and
$\Omega_m = 0.3163 \pm 0.019$.
The equation-of-state parameters are constrained to
$w_0 = -0.93 \pm 0.24$ and $w_a = -0.57 \pm 0.21$.
The complementary nature of BAO and supernova observations substantially improves the parameter constraints while remaining fully consistent with a cosmological-constant interpretation at the $1\sigma$ level. The inferred value of $H_0$ corresponds to tensions of $2.01\sigma$ with Planck and $1.45\sigma$ with SH0ES.

\paragraph{CMB + CC + Pantheon constraints.}
The joint CMB, CC, and Pantheon analysis prefers a higher Hubble constant,
$H_0 = 71.76 \pm 1.70~\mathrm{km\,s^{-1}\,Mpc^{-1}}$, with
$\Omega_m = 0.302 \pm 0.035$.
The CPL parameters are constrained to
$w_0 = -0.85 \pm 0.21$ and $w_a = -0.73 \pm 0.20$.
This combination remains compatible with mildly evolving dark energy, although no statistically significant deviation from $\Lambda$CDM is detected. The higher value of $H_0$ leads to the largest remaining discrepancy with Planck ($2.45\sigma$) while reducing the SH0ES tension to only $0.66\sigma$.

\paragraph{Full combination: CMB + CC + BAO + Pantheon.}
Finally, combining all datasets yields
$H_0 = 70.21 \pm 1.44~\mathrm{km\,s^{-1}\,Mpc^{-1}}$ and
$\Omega_m = 0.302 \pm 0.012$.
The CPL parameters are constrained to
$w_0 = -0.83 \pm 0.16$ and $w_a = -0.79 \pm 0.14$.
The combined dataset provides the tightest constraints on the cosmological parameters while remaining statistically consistent with a mildly evolving dark energy scenario. Although the full combination partially alleviates the Hubble tension, with
$T_{\rm Planck}=1.87\sigma$ and $T_{\rm R22}=1.68\sigma$, it does not completely resolve the discrepancy, suggesting that late-time modifications to the dark energy sector alone are insufficient to fully reconcile early- and late-Universe measurements. All results are summarized in Table~\ref{tab:cpl_summary} and are in broad agreement with recent determinations of $H_0$ \cite{Y1,Y2,Y3,Y4}. Recent analyses of the CPL parameters using DESI DR2, Pantheon+, and CMB data likewise indicate values compatible with mildly dynamical dark energy and $H_0$ in the range $69$--$71~\mathrm{km\,s^{-1}\,Mpc^{-1}}$ \cite{DESIDR2_2025,LiuLiWang2025,Malekjani2025,Park2024}.

\begin{figure}[H]
	\includegraphics[width=\columnwidth]{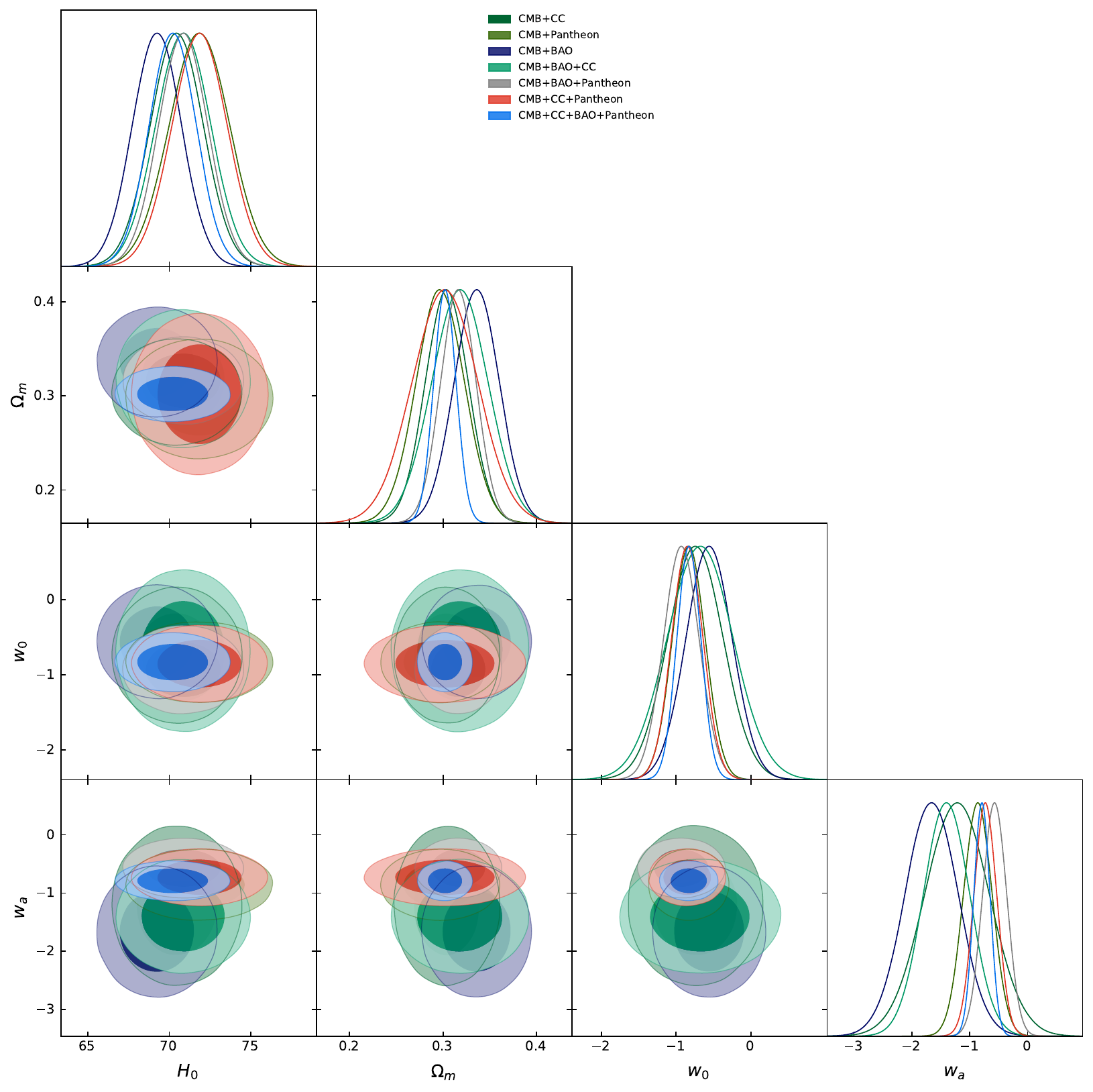}
	\caption{Posterior distributions of the CPL dark energy parameters ($w_0$, $w_a$) and the Hubble constant $H_0$ derived from the Bayesian Physics-Informed Neural Network (BPINN) analysis using various low-redshift datasets. The filled contours correspond to the $68\%$ and $95\%$ confidence levels, illustrating the parameter correlations and dataset-specific constraints.}
	\label{fig:cpl}
\end{figure}

\begin{table}[H]
	\centering
	\caption{Summary of CPL parameter constraints and the corresponding Hubble tension levels with respect to Planck 2018 and SH0ES (R22). All uncertainties are quoted at the $1\sigma$ confidence level.}
	\label{tab:cpl_summary}
	\begin{tabular}{lcccccc}
		\hline
		Dataset
		& $H_0$ [km\,s$^{-1}$\,Mpc$^{-1}$]
		& $\Omega_m$
		& $w_0$
		& $w_a$
		& $T_{\rm Planck}$ [$\sigma$]
		& $T_{\rm R22}$ [$\sigma$] \\
		\hline
		
		CMB+CC
		& $70.46 \pm 1.63$
		& $0.304 \pm 0.023$
		& $-0.74 \pm 0.37$
		& $-1.22 \pm 0.56$
		& $1.84$
		& $1.39$ \\
		
		CMB+Pantheon
		& $71.84 \pm 1.85$
		& $0.297 \pm 0.026$
		& $-0.83 \pm 0.22$
		& $-0.86 \pm 0.25$
		& $2.34$
		& $0.56$ \\
		
		CMB+BAO
		& $69.24 \pm 1.51$
		& $0.336 \pm 0.024$
		& $-0.56 \pm 0.31$
		& $-1.66 \pm 0.46$
		& $1.16$
		& $2.02$ \\
		
		CMB+BAO+CC
		& $70.81 \pm 1.68$
		& $0.318 \pm 0.030$
		& $-0.68 \pm 0.44$
		& $-1.40 \pm 0.40$
		& $1.97$
		& $1.19$ \\
		
		CMB+BAO+Pantheon
		& $70.54 \pm 1.51$
		& $0.3163 \pm 0.019$
		& $-0.93 \pm 0.24$
		& $-0.57 \pm 0.21$
		& $2.01$
		& $1.45$ \\
		
		CMB+CC+Pantheon
		& $71.76 \pm 1.70$
		& $0.302 \pm 0.035$
		& $-0.85 \pm 0.21$
		& $-0.73 \pm 0.20$
		& $2.45$
		& $0.66$ \\
		
		CMB+CC+BAO+Pantheon
		& $70.21 \pm 1.44$
		& $0.302 \pm 0.012$
		& $-0.83 \pm 0.16$
		& $-0.79 \pm 0.14$
		& $1.87$
		& $1.68$ \\
		
		\hline
	\end{tabular}
\end{table}
The reconstructed posterior distributions for the CPL parameters and $H_0$ are shown in Figure~\ref{fig:cpl}, which compares the impact of different datasets on the inferred dark energy dynamics.

\begin{figure}[H]
	\includegraphics[width=\columnwidth]{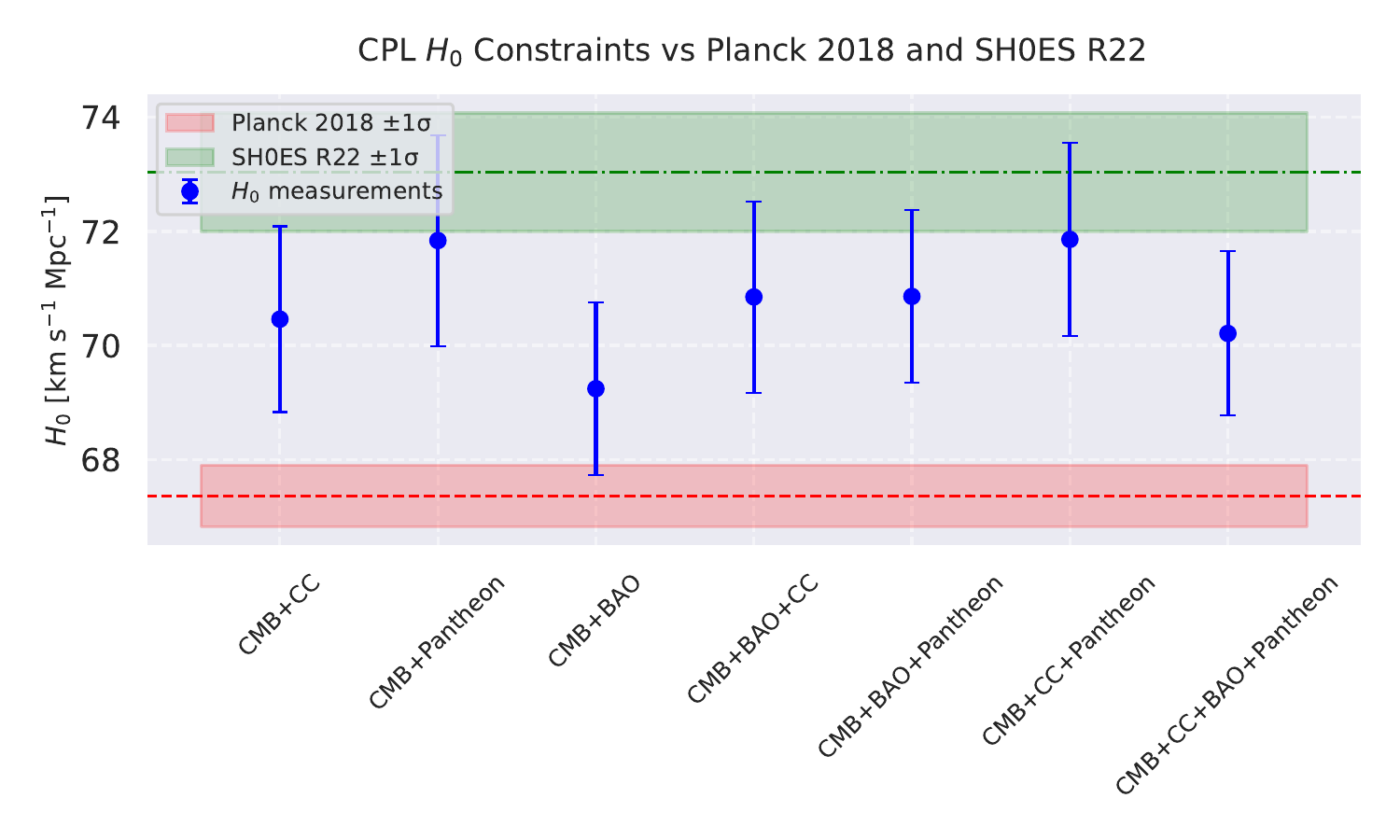}
	\caption{Comparison the Hubble constant $H_0$ derived from the Bayesian Physics-Informed Neural Network (BPINN) analysis using various low-redshift datasets with Planck 2018 and R22.}
	\label{fig:h0cpl}
\end{figure}

The inferred values of the Hubble constant $H_0$ from different low-redshift datasets are illustrated in Figure~\ref{fig:h0cpl}, showing a comparison with the Planck 2018 and R22 measurements.
\subsection*{Physical Interpretation of the CPL Dark Energy Constraints}

The CPL parametrization provides a convenient phenomenological framework for investigating possible departures from a cosmological constant while remaining sufficiently flexible to capture smooth late-time evolution of the dark-energy equation of state. The Bayesian constraints obtained in this work consistently favor present-day values in the range
$-0.93 \lesssim w_0 \lesssim -0.56$,
whereas the evolution parameter remains negative,
$-1.66 \lesssim w_a \lesssim -0.57$,
for all dataset combinations. Although the best-fit values deviate moderately from the $\Lambda$CDM prediction, the associated uncertainties indicate that a cosmological constant remains fully compatible with the observations at approximately the $1\sigma$ confidence level.

\paragraph{Present-day equation of state.}

The inferred values of $w_0$ are consistently larger than $-1$, indicating that the preferred cosmological solutions lie within the quintessence region rather than the phantom regime. Nevertheless, the statistical uncertainties remain sufficiently broad that the cosmological-constant limit ($w_0=-1$) is included within the allowed parameter space for every dataset combination. Consequently, the present analysis does not provide statistically significant evidence for either phantom dark energy or a fundamental departure from $\Lambda$CDM. Instead, the reconstructed equation of state should be interpreted as indicating, at most, a mild preference for an evolving dark-energy component whose present-day behavior remains close to vacuum energy.

\paragraph{Evolution of the dark-energy equation of state.}

A notable feature of the present analysis is the consistently negative value of the evolution parameter $w_a$. Within the CPL parametrization,
\[
w(z)=w_0+w_a\frac{z}{1+z},
\]
negative values of $w_a$ imply that the equation of state becomes progressively more negative with increasing redshift. However, because both $w_0$ and $w_a$ possess relatively large uncertainties, the data do not require strong time evolution. Instead, the observations favor a slowly varying equation of state whose evolution remains fully consistent with a cosmological constant within the current observational precision. Such gradual evolution is phenomenologically compatible with a broad class of quintessence models as well as effective descriptions emerging from modified gravity or interacting dark-sector scenarios \cite{NojiriOdintsov2006,NojiriOdintsov2011,Nojiri2025,Nojiri2026}.

\paragraph{Implications for cosmic expansion.}

The reconstructed CPL parameters naturally modify the late-time expansion history without introducing large departures from the standard cosmological model. Compared with $\Lambda$CDM, the inferred expansion history allows moderately larger values of the Hubble constant, with the preferred range extending from approximately
$69.2$ to $71.8~\mathrm{km\,s^{-1}\,Mpc^{-1}}$
depending on the adopted observational combination. As a consequence, the tension with the SH0ES local measurement is systematically reduced, while a moderate discrepancy with the \textit{Planck} 2018 inference remains. This behavior indicates that smooth late-time dark-energy evolution can partially reconcile early- and late-Universe determinations of $H_0$, although it does not completely eliminate the existing disagreement.

\paragraph{Physical implications and theoretical consistency.}

Overall, the Bayesian reconstruction supports a cosmological scenario that remains close to $\Lambda$CDM while allowing modest departures through a slowly evolving dark-energy component. Since the preferred solutions remain within the quintessence region and do not require $w_0<-1$, the present results avoid the theoretical difficulties commonly associated with phantom fields, including ghost and gradient instabilities \cite{Vikman2005}. At the same time, the reconstructed evolution remains sufficiently flexible to accommodate effective descriptions arising in scalar-tensor theories, modified gravity models, or interacting dark-sector scenarios \cite{NojiriOdintsov2006,NojiriOdintsov2011,Nojiri2025,Nojiri2026}. Although the CPL extension substantially improves the consistency between independent cosmological probes, the persistence of residual tension between early- and late-Universe measurements suggests that late-time dark-energy dynamics alone are unlikely to provide a complete solution to the Hubble tension. Additional physical ingredients, such as early dark energy, neutrino physics, or viable modifications of gravity, may therefore still be required to achieve a fully consistent cosmological framework \cite{HuSawicki2007}.

\subsection{Bayesian Constraints on CPL Dark Energy with Massive Neutrinos}

Within the Bayesian Physics-Informed Neural Network (BPINN) framework, the CMB distance priors
are incorporated as an additional Gaussian likelihood term in the total loss function, so that the
early-Universe geometry and the late-time expansion history are fit jointly within the same
posterior. Following the treatment described in Sec.~IV.D, the decoupling redshift $z_*$ and the
sound horizon $r_s(z_*)$ entering the shift parameter $R$ and the acoustic scale $\ell_A$ are
evaluated with the standard early-Universe fitting formulas of Eqs.~(43)--(44), using the sampled
values of $\omega_b$ and $\omega_m$ at each step; they are not obtained by extrapolating the
PINN-reconstructed $H(z)$, which is trained and evaluated only over the low-redshift range covered
by the CC, BAO, and Pantheon$^{+}$ data. Since the neutrino density parameter,
$\Omega_\nu \propto \Sigma m_\nu$, contributes to $\omega_m$, the CMB likelihood still responds to
the sampled neutrino mass through this early-Universe channel, in parallel with its effect on the
late-time background evolution entering the BAO, CC, and supernova likelihoods. This coupling lets
the joint fit constrain $\Sigma m_\nu$, $H_0$, and the CPL parameters $(w_0, w_a)$ using both
early- and late-Universe information, without requiring the late-time PINN to hold outside its
intended redshift range.

The CPL parametrization extended by a free summed neutrino mass, $\Sigma m_\nu$, represents one of the simplest phenomenological extensions of $\Lambda$CDM capable of simultaneously modifying the cosmic expansion history and the matter clustering amplitude. Table~\ref{tab:cpl_mnu_summary} summarizes the inferred cosmological parameters obtained from different combinations of CMB distance priors, Cosmic Chronometers (CC), Baryon Acoustic Oscillations (BAO), and Pantheon Type Ia supernovae. In addition to the cosmological parameters, the table reports the statistical tension of the inferred Hubble constant relative to both the \textit{Planck} 2018 CMB determination and the SH0ES (R22) local measurement.

\subsection*{Late-Time Expansion and Dynamical Dark Energy}

Unlike previous CPL analyses without massive neutrinos, the present results show that allowing a free summed neutrino mass shifts the preferred dark-energy parameter space toward values that are generally consistent with a cosmological constant within approximately the $1\sigma$ confidence interval. The inferred present-day equation-of-state parameter spans
$-0.96 \lesssim w_0 \lesssim -0.55$,
depending on the adopted data combination, with the strongest preference for values close to $w_0=-1$ arising from the combinations including Pantheon data. In particular, the CMB+CC+Pantheon dataset yields
$w_0=-0.96\pm0.21$,
while the full CMB+CC+BAO+Pantheon combination gives
$w_0=-0.801\pm0.180$,
both remaining statistically compatible with $\Lambda$CDM.

The time-evolution parameter exhibits a wider variation among the different datasets,
ranging from
$w_a=-1.44\pm0.24$
for the CMB+BAO combination to
$w_a=-0.61\pm0.19$
for the full data combination.
Although negative values of $w_a$ are mildly favored, no statistically significant evidence for rapid evolution of the dark-energy equation of state is found. Overall, the CPL sector remains well constrained despite the inclusion of an additional neutrino-mass degree of freedom, indicating that the combined late-time datasets effectively break the principal parameter degeneracies.

\subsection*{Constraints on the Summed Neutrino Mass}

The summed neutrino mass remains tightly constrained by all observational combinations. The inferred $2\sigma$ upper limits range from
$\Sigma m_\nu<0.28~\mathrm{eV}$
for CMB+Pantheon
to
$\Sigma m_\nu<0.16~\mathrm{eV}$
for the full CMB+CC+BAO+Pantheon dataset.
These constraints are consistent with recent cosmological analyses employing compressed CMB likelihoods and DESI-era late-time observations \cite{Y5,Y6}.

The progressive tightening of the neutrino-mass bound as complementary datasets are added demonstrates the complementarity between geometric probes and direct measurements of the expansion history. Although no statistically significant detection of non-zero neutrino mass is obtained, the allowed parameter space remains sufficiently broad for neutrinos to influence the inferred expansion history through their correlations with $H_0$ and the dark-energy sector.

\subsection*{Implications for the Hubble Tension}

The inferred values of the Hubble constant remain stable across all dataset combinations, spanning the interval
$69.73 \lesssim H_0 \lesssim 71.63~\mathrm{km\,s^{-1}\,Mpc^{-1}}$.
In particular, every dataset combination reduces the discrepancy with the SH0ES (R22) determination to below the $2\sigma$ level. The most significant improvement is obtained for the CMB+CC+Pantheon combination, for which the residual tension decreases to only
$0.83\sigma$,
whereas the remaining combinations lie between approximately
$1.18\sigma$
and
$1.83\sigma$.
This represents a substantial improvement compared with the standard $\Lambda$CDM scenario.

On the other hand, a moderate disagreement with the \textit{Planck} 2018 inference persists. Depending on the adopted dataset combination, the residual tension ranges from approximately
$1.6\sigma$
to
$2.8\sigma$,
with the largest value corresponding to the CMB+CC+Pantheon combination.
This asymmetric behaviour indicates that the CPL+$\Sigma m_\nu$ extension preferentially shifts the cosmological solution toward the late-Universe determination of the Hubble constant while remaining broadly compatible with early-Universe constraints.

\subsection*{Physical Interpretation}

The combined Bayesian analysis indicates that introducing massive neutrinos into the CPL framework produces a statistically stable extension of the standard cosmological model without generating strong degeneracies in the reconstructed parameter space. The preferred solutions remain close to the $\Lambda$CDM limit, while allowing moderate departures in both $w_0$ and $w_a$ that improve the overall consistency among independent cosmological probes.

Although the present analysis does not provide evidence for a non-zero neutrino mass detection or a definitive departure from a cosmological constant, it demonstrates that the combined effects of dynamical dark energy and massive neutrinos significantly alleviate the Hubble tension with local measurements while preserving consistency with CMB observations. Future high-precision measurements from next-generation BAO surveys, improved cosmic chronometer compilations, and forthcoming Type Ia supernova datasets are expected to further tighten the allowed CPL parameter space and help clarify whether these mild deviations from $\Lambda$CDM reflect new late-time physics or statistical fluctuations.

\begin{figure}[H]
	\includegraphics[width=\columnwidth]{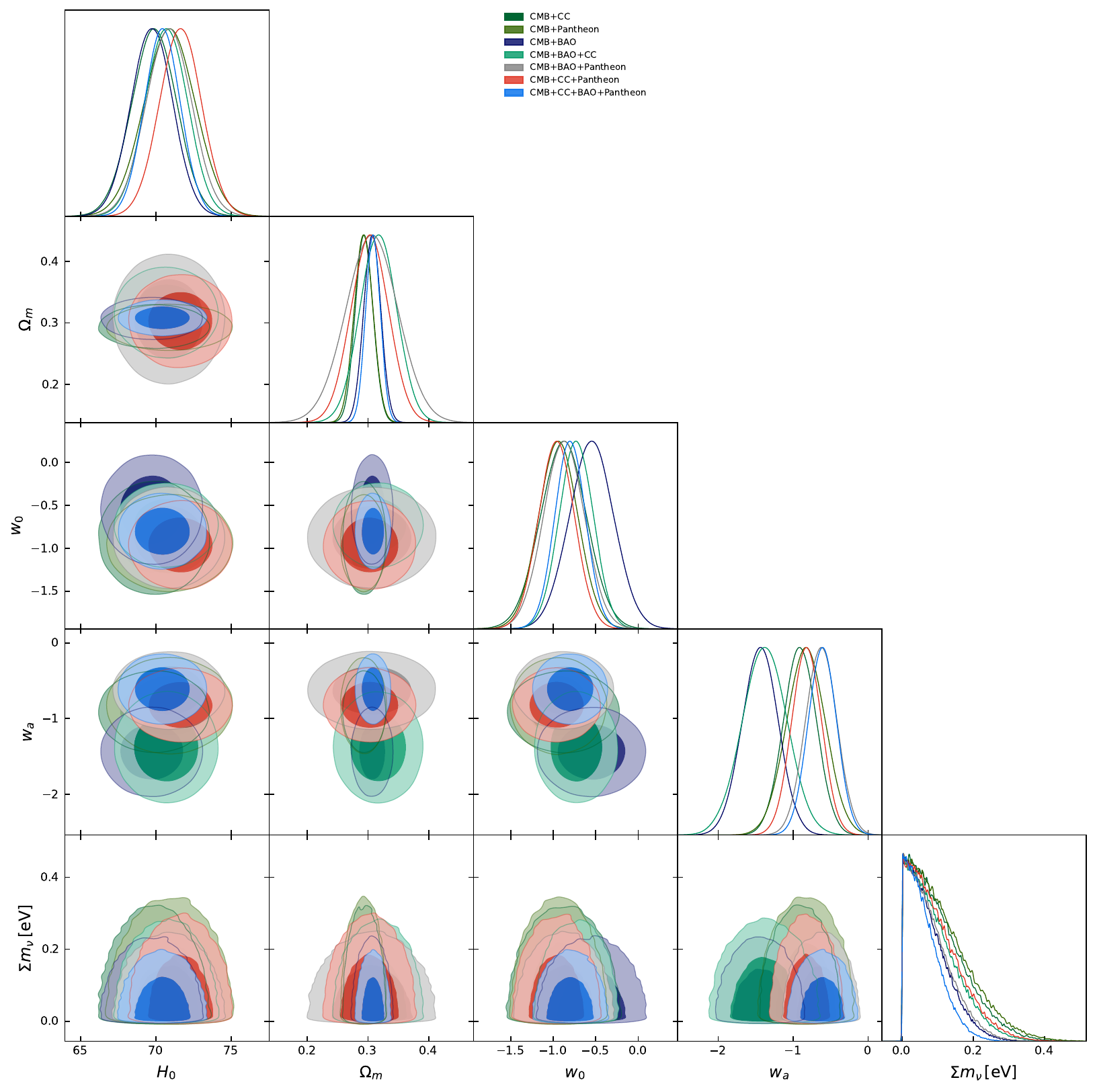}
	\caption{Corner plot of the CPL+$\Sigma m_\nu$ posterior distributions obtained from the Bayesian Physics-Informed Neural Network (BPINN) analysis. The filled contours correspond to the $68\%$ and $95\%$ confidence levels, showing the correlations between $H_0$, $\Omega_m$, $w_0$, $w_a$, and the sum of neutrino masses $\Sigma m_\nu$.}
	
	\label{fig:cpl2}
\end{figure}

\begin{table}[H]
	\centering
	\caption{Summary of CPL+$\Sigma m_\nu$ parameter constraints and the corresponding Hubble tension levels with respect to Planck 2018 and SH0ES (R22). All uncertainties are quoted at the $1\sigma$ confidence level. Upper bounds on $\Sigma m_\nu$ correspond to the $2\sigma$ limits.}
	\label{tab:cpl_mnu_summary}
	\begin{tabular}{lcccccc}
		\hline
		Dataset
		& $H_0$ [km\,s$^{-1}$\,Mpc$^{-1}$]
		& $\Omega_m$
		& $w_0$
		& $w_a$
		& $\Sigma m_\nu$ [eV]
		& $T_{\rm Planck}$ / $T_{\rm R22}$ [$\sigma$] \\
		\hline
		
		CMB+CC
		& $69.88 \pm 1.50$
		& $0.2937 \pm 0.0145$
		& $-0.88 \pm 0.27$
		& $-0.91 \pm 0.22$
		& $<0.26$
		& $1.60 \,/\, 1.83$ \\
		
		CMB+Pantheon
		& $70.91 \pm 1.70$
		& $0.2927 \pm 0.0155$
		& $-0.940 \pm 0.23$
		& $-0.83 \pm 0.26$
		& $<0.28$
		& $2.17 \,/\, 1.29$ \\
		
		CMB+BAO
		& $69.73 \pm 1.38$
		& $0.3070 \pm 0.0140$
		& $-0.55 \pm 0.26$
		& $-1.44 \pm 0.24$
		& $<0.19$
		& $1.65 \,/\, 1.82$ \\
		
		CMB+BAO+CC
		& $70.69 \pm 1.41$
		& $0.3169 \pm 0.0302$
		& $-0.729 \pm 0.20$
		& $-1.380 \pm 0.30$
		& $<0.21$
		& $2.22 \,/\, 1.72$ \\
		
		CMB+BAO+Pantheon
		& $70.86 \pm 1.51$
		& $0.3063 \pm 0.0429$
		& $-0.873 \pm 0.24$
		& $-0.627 \pm 0.21$
		& $<0.20$
		& $2.28 \,/\, 1.18$ \\
		
		CMB+CC+Pantheon
		& $71.63 \pm 1.40$
		& $0.3030 \pm 0.0310$
		& $-0.960 \pm 0.210$
		& $-0.8200 \pm 0.20$
		& $<0.24$
		& $2.79 \,/\, 0.83$ \\
		
		CMB+CC+BAO+Pantheon
		& $70.41 \pm 1.28$
		& $0.3082 \pm 0.0120$
		& $-0.801 \pm 0.180$
		& $-0.6100 \pm 0.19$
		& $<0.16$
		& $2.5 \,/\, 1.51$ \\
		
		\hline
	\end{tabular}
\end{table}
The combined constraints on the CPL parameters and the neutrino mass are presented in Figure~\ref{fig:cpl2}, which shows the posterior distributions and their mutual correlations.

\begin{figure}[H]
	\includegraphics[width=\columnwidth]{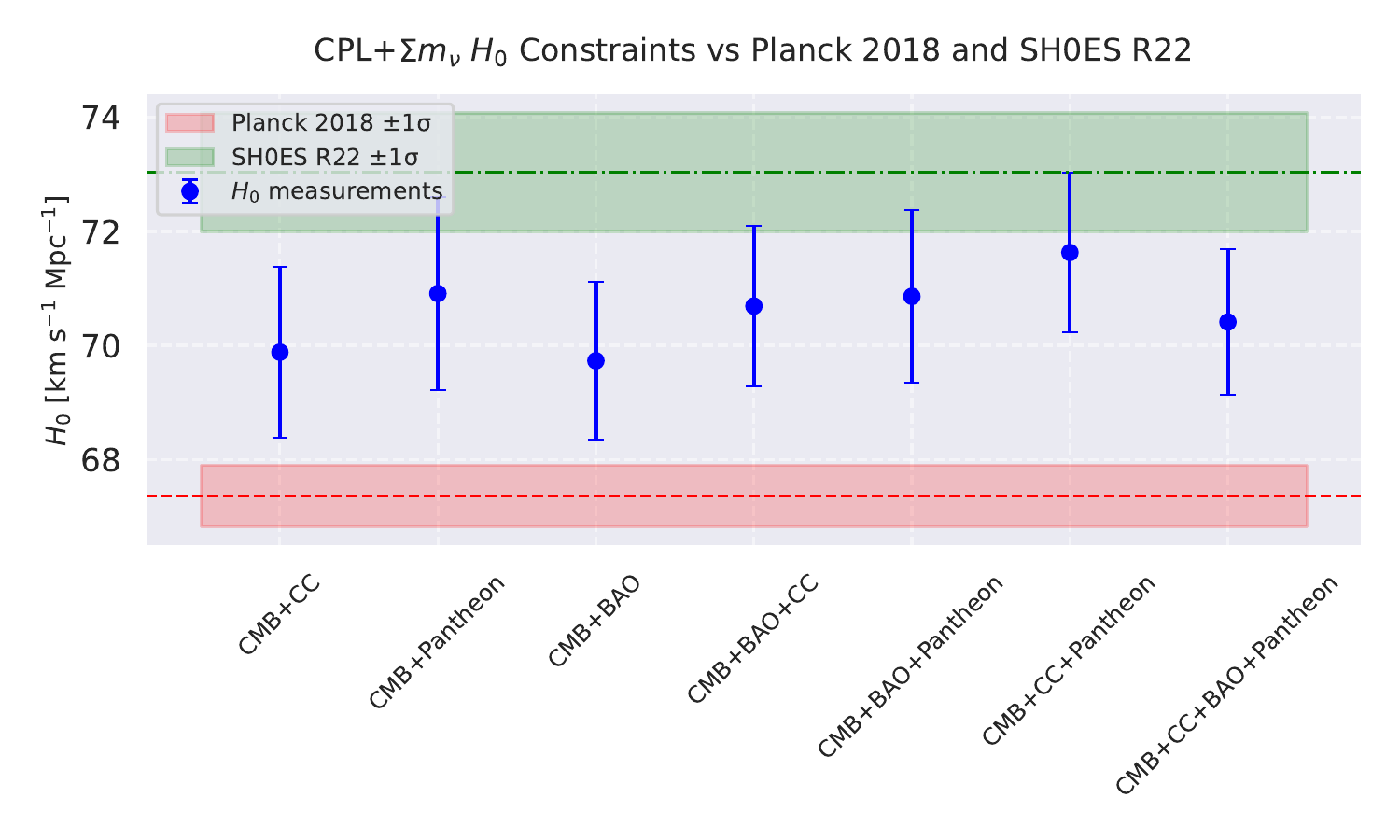}
	\caption{Comparison the Hubble constant $H_0$ derived from the Bayesian Physics-Informed Neural Network (BPINN) analysis using various low-redshift datasets with Planck 2018 and R22.}
	\label{fig:cplh0mnu}
\end{figure}

The posterior estimates of the Hubble constant $H_0$ obtained from the BPINN analysis incorporating massive neutrinos are presented in Figure~\ref{fig:cplh0mnu}, providing a comparison with Planck 2018 and R22 measurements.

\section{Bayesian Physics-Informed Neural Network Constraints on the wCDM Model}
\label{sec:bpinn_lcdm}

In this section, we present the cosmological constraints obtained for the standard wCDM model using a Bayesian Physics-Informed Neural Network (BPINN) framework. The analysis combines different low-redshift observational probes, and the resulting parameter constraints are summarized in Table~\ref{tab:lcdm_summary}. For each dataset and data combination, we discuss the inferred values of the Hubble constant $H_0$, the matter density parameter $\Omega_m$, and the dark energy equation-of-state parameter $w_0$, as well as the corresponding levels of Hubble tension with respect to Planck~2018 and SH0ES (R22).

\paragraph{Cosmic Chronometers (CC).}
Using Cosmic Chronometer data alone, we obtain $H_0 = 70.06 \pm 1.43~\mathrm{km\,s^{-1}\,Mpc^{-1}}$, which lies between the Planck~2018 and SH0ES determinations. This results in a moderate tension of $T_{\rm Planck} \simeq 1.8\sigma$ and $T_{\rm R22} \simeq 1.6\sigma$. The inferred matter density $\Omega_m = 0.312 \pm 0.014$ and equation-of-state parameter $w_0 = -1.011 \pm 0.027$ are fully consistent with a cosmological constant. These results indicate that CC data mildly favor a higher late-time expansion rate compared to the CMB, but are insufficient on their own to resolve the Hubble tension.

\paragraph{Pantheon Supernovae.}
The Pantheon Type~Ia supernova sample prefers a higher value of the Hubble constant, $H_0 = 71.22 \pm 1.72$, leading to an increased tension with Planck at the level of $T_{\rm Planck} \simeq 2.1\sigma$, while substantially reducing the discrepancy with SH0ES to about $1.0\sigma$. The values of $\Omega_m$ and $w_0$ remain consistent with $\Lambda$CDM expectations. This behavior reflects the well-known tendency of distance-ladder measurements to favor a faster late-time expansion compared to early-Universe probes.

\paragraph{Baryon Acoustic Oscillations (BAO).}
In contrast, BAO data alone yield a lower Hubble constant, $H_0 = 69.25 \pm 1.33$, which is more compatible with Planck~2018, resulting in a reduced tension of $T_{\rm Planck} \simeq 1.4\sigma$. However, this comes at the expense of an increased tension with SH0ES, reaching $T_{\rm R22} \simeq 2.3\sigma$. The tight constraint on $\Omega_m = 0.319 \pm 0.004$ reflects the strong geometrical nature of BAO measurements and their sensitivity to the early-Universe calibration of the sound horizon.

\paragraph{BAO+CC Combination.}
When BAO data are combined with Cosmic Chronometers, the resulting constraint shifts to $H_0 = 68.45 \pm 1.55$, further improving consistency with Planck ($T_{\rm Planck} \simeq 0.6\sigma$) while significantly increasing the tension with SH0ES to $T_{\rm R22} \simeq 2.5\sigma$. This indicates that the inclusion of CC data does not overcome the intrinsic preference of BAO for lower values of $H_0$, reinforcing the early-Universe anchored expansion history.

\paragraph{BAO+Pantheon Combination.}
The joint analysis of BAO and Pantheon data yields an intermediate value of the Hubble constant, $H_0 = 70.44 \pm 1.40$. In this case, both tensions remain at the level of $1.5$--$2\sigma$, suggesting a partial statistical compromise between early- and late-time probes. The inferred parameters remain consistent with wCDM, but the Hubble tension is not fully alleviated.

\paragraph{CC+Pantheon Combination.}
Combining the two late-time probes, CC and Pantheon, leads to a higher expansion rate, $H_0 = 70.64 \pm 1.65$, which increases the tension with Planck to $T_{\rm Planck} \simeq 2.1\sigma$ while reducing the discrepancy with SH0ES to approximately $1.3\sigma$. This result suggests that late-time data alone tend to favor higher values of $H_0$, yet remain insufficient to reconcile all measurements within the $\Lambda$CDM framework.

\begin{figure}[H]
	\includegraphics[width=\columnwidth]{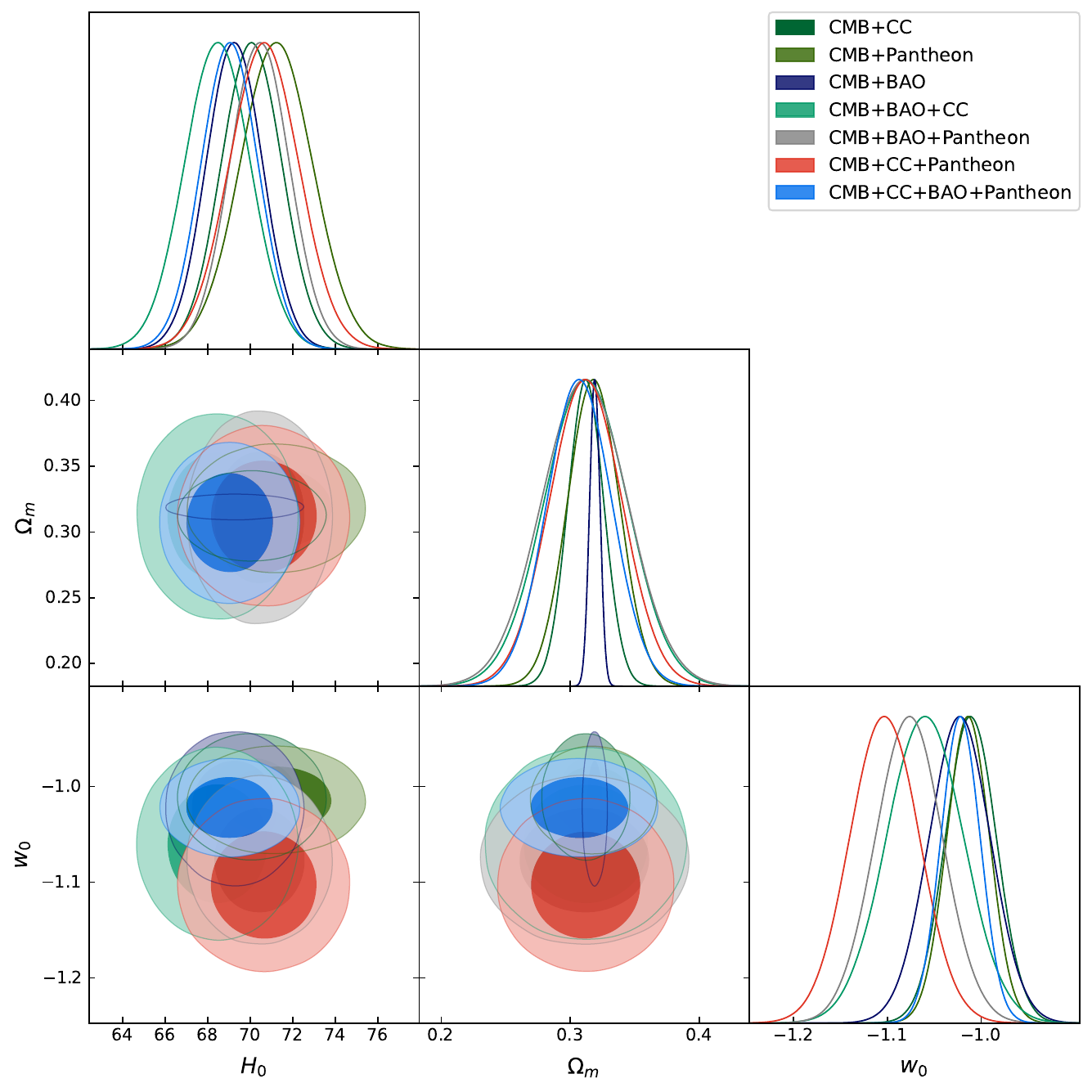}
	\caption{Corner plot of the wCDM posterior distributions obtained from the Bayesian Physics-Informed Neural Network (BPINN) analysis for different low-redshift datasets. The filled contours correspond to the $68\%$ and $95\%$ confidence levels. Each color represents a different dataset or combination of datasets as indicated in the legend.}
	\label{fig:cpl3}
\end{figure}

\paragraph{CC+BAO+Pantheon Combination.}
Finally, the full combination of CC, BAO, and Pantheon data yields $H_0 = 69.03 \pm 1.35$, representing a balanced compromise between early- and late-time measurements. Nevertheless, significant residual tensions persist with both Planck ($T_{\rm Planck} \simeq 1.2\sigma$) and SH0ES ($T_{\rm R22} \simeq 2.1\sigma$). This outcome demonstrates that even within a Bayesian PINN framework and using combined low-redshift datasets, the $\Lambda$CDM model is unable to fully resolve the Hubble tension, thereby motivating the exploration of extensions beyond the standard cosmological paradigm. The posterior distributions of the $\Lambda$CDM parameters for each dataset are shown in the corner plot in Figure~\ref{fig:cpl3}, which shows the constraints on $H_0$, $\Omega_m$, and $w_0$ and their correlations.

\begin{figure}[H]
	\includegraphics[width=\columnwidth]{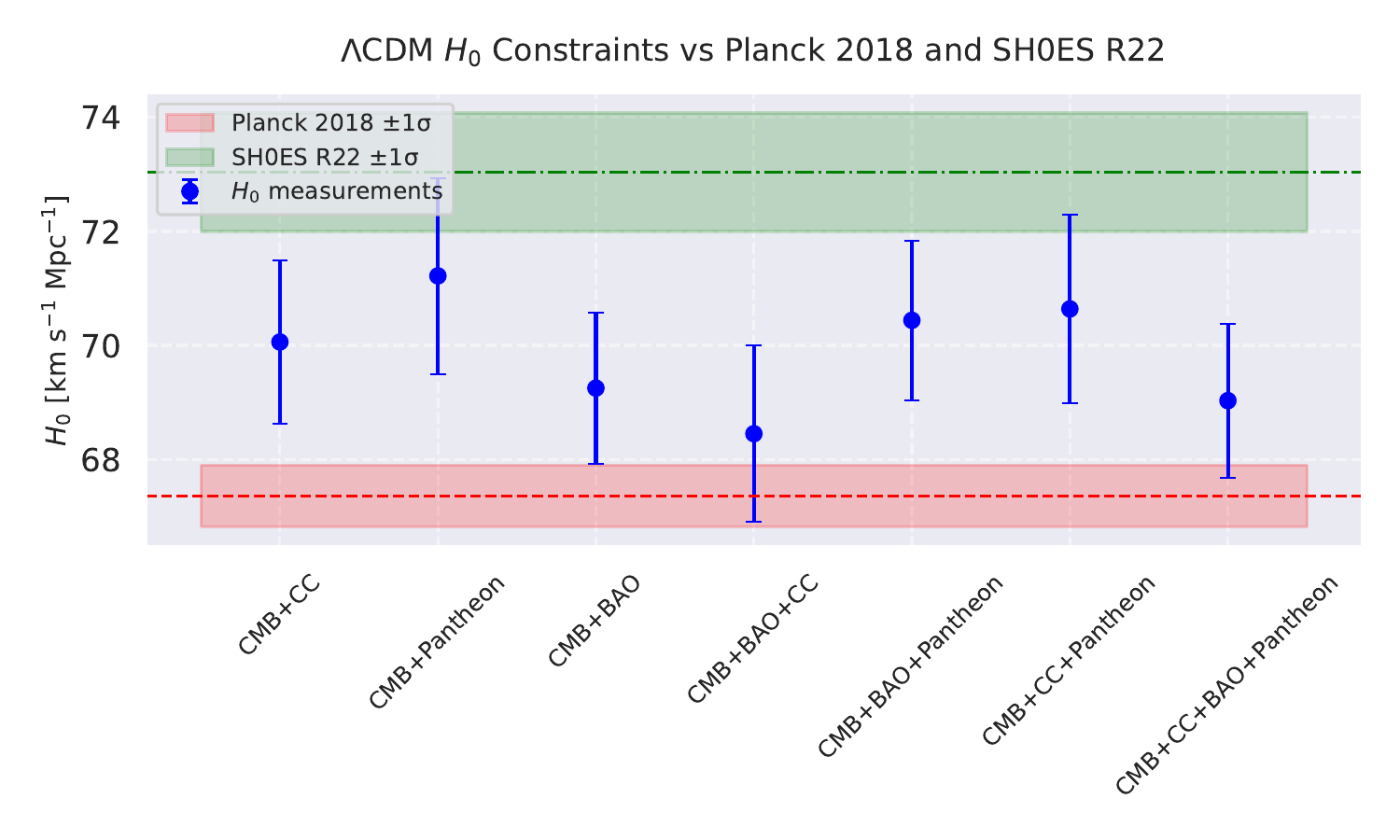}
	\caption{Comparison the Hubble constant $H_0$ derived from the Bayesian Physics-Informed Neural Network (BPINN) analysis using various low-redshift datasets with Planck 2018 and R22.}
	\label{fig:cpllcdm}
\end{figure}

Figure~\ref{fig:cpllcdm} shows the Hubble constant $H_0$ inferred from the BPINN analysis under the wCDM model using various low-redshift datasets, compared with the Planck 2018 and R22 measurements.

\begin{table}[H]
	\centering
	\caption{Summary of wCDM parameter constraints and the corresponding Hubble tension levels with respect to Planck 2018 and SH0ES (R22). All uncertainties are quoted at the $1\sigma$ confidence level.}
	\label{tab:lcdm_summary}
	\begin{tabular}{lccccc}
		\hline
		Dataset 
		& $H_0$ [km\,s$^{-1}$\,Mpc$^{-1}$] 
		& $\Omega_m$ 
		& $w_0$ 
		& $T_{\rm Planck}$ [$\sigma$] 
		& $T_{\rm R22}$ [$\sigma$] \\
		\hline
		CMB+CC 
		& $70.06 \pm 1.43$ 
		& $0.312 \pm 0.014$ 
		& $-1.011 \pm 0.027$ 
		& $1.78$ 
		& $1.56$ \\
		
		CMB+Pantheon 
		& $71.22 \pm 1.72$ 
		& $0.318 \pm 0.020$ 
		& $-1.014 \pm 0.023$ 
		& $2.08$ 
		& $1.04$ \\
		
		CMB+BAO 
		& $69.25 \pm 1.33$ 
		& $0.319 \pm 0.004$ 
		& $-1.023 \pm 0.033$ 
		& $1.36$ 
		& $2.29$ \\
		
		CMB+BAO+CC 
		& $68.45 \pm 1.55$ 
		& $0.312 \pm 0.032$ 
		& $-1.060 \pm 0.041$ 
		& $0.63$ 
		& $2.48$ \\
		
		CMB+BAO+Pantheon 
		& $70.44 \pm 1.40$ 
		& $0.311 \pm 0.033$ 
		& $-1.077 \pm 0.036$ 
		& $2.01$ 
		& $1.45$ \\
		
		CMB+CC+Pantheon 
		& $70.64 \pm 1.65$ 
		& $0.312 \pm 0.028$ 
		& $-1.103 \pm 0.037$ 
		& $2.05$ 
		& $1.29$ \\
		
		CMB+CC+BAO+Pantheon 
		& $69.03 \pm 1.35$ 
		& $0.307 \pm 0.025$ 
		& $-1.022 \pm 0.021$ 
		& $1.21$ 
		& $2.14$ \\
		\hline
	\end{tabular}
\end{table}
\section{Comparative Analysis of wCDM, CPL, and CPL+$\Sigma m_\nu$ Models within the Bayesian PINN Framework}

We compare the inferred Hubble constant $H_0$ and the resulting Hubble-tension levels obtained under the wCDM, CPL, and CPL+$\Sigma m_\nu$ models (Tables~\ref{tab:lcdm_summary}--\ref{tab:cpl_summary}), all derived within the same Bayesian PINN pipeline, so that differences reflect model content rather than analysis inconsistencies.

\subsection{$H_0$ and the Hubble tension}

For wCDM (Table~\ref{tab:lcdm_summary}), $H_0$ spans $68.5$--$71.2\ \mathrm{km\,s^{-1}\,Mpc^{-1}}$: BAO-inclusive combinations pull $H_0$ downward, minimizing the Planck tension (CMB+BAO+CC: $T_{\rm Planck}=0.63\sigma$) at the cost of a larger R22 tension ($T_{\rm R22}=2.48\sigma$), reflecting the known anti-correlation between the two tensions across dataset combinations.

Allowing $w_0,w_a$ to vary (CPL, Table~\ref{tab:cpl_summary}) shifts $H_0$ systematically upward ($H_0\gtrsim70$, up to $71.84\pm1.85$ for CMB+Pantheon), with mildly quintessence-like  evolving equations of state. This reduces $T_{\rm R22}$ substantially (down to $0.56\sigma$ for CMB+Pantheon) while increasing $T_{\rm Planck}$, indicating that dynamical dark energy alone can ease the late-time tension without fully resolving the early--late discrepancy.

Adding a neutrino mass component (CPL+$\Sigma m_\nu$, Table~II) stabilizes $H_0$ in a
narrower range ($69.7$--$71.6$) with $\Sigma m_\nu \lesssim 0.16$--$0.28\,\mathrm{eV}$. The
effect on the balance between the two tensions is not uniform: for four of the seven
combinations (CMB+CC, CMB+Pantheon, CMB+BAO, CMB+BAO+CC) the gap between $T_{\rm Planck}$
and $T_{\rm R22}$ narrows once $\Sigma m_\nu$ is included, while for the remaining three --
including the full CMB+CC+BAO+Pantheon combination ($T_{\rm Planck}=2.5\sigma$,
$T_{\rm R22}=1.51\sigma$, a larger gap than the $1.87\sigma/1.68\sigma$ found without
$\Sigma m_\nu$) -- it widens instead. Rather than systematically balancing the two tensions,
the coupling between dark-energy dynamics and $\Sigma m_\nu$ appears to redistribute them in
a way that depends on which datasets are combined.

The Bayesian PINN framework underlies all three fits: embedding the field equations in the loss enforces physical consistency while enabling data-driven reconstruction of $H(z)$, and its Bayesian formulation propagates epistemic uncertainty in a principled way — an advantage for extended models prone to parameter degeneracies under standard parametric fitting.

An interesting difference between the two parametrizations concerns the sign of $w_0-(-1)$.
Every wCDM fit in Table~III returns $w_0$ slightly below $-1$, and for two combinations --
CMB+CC+Pantheon ($w_0=-1.103\pm0.037$) and CMB+BAO+Pantheon ($w_0=-1.077\pm0.036$) -- this
reaches $2.1$--$2.8\sigma$, a mild but not entirely negligible pull into the phantom regime.
None of the CPL fits in Table~I show anything similar: once $w_a$ is free to vary, $w_0$
settles above $-1$ in every combination, including these same two. We read this as a
consequence of the parametrization rather than a change in what the data prefer. A
constant-$w$ model has only one number to describe the equation of state across the full
redshift range probed jointly by CC, BAO, and Pantheon$^{+}$; if these probes pull, even
mildly, toward different effective values of $w(z)$ at their respective redshifts, $w_0$ can
only settle at a compromise, and that compromise may fall below $-1$ without any single
dataset actually requiring phantom dark energy. CPL removes this constraint by letting $w_a$
carry the redshift dependence instead, consistent with the negative $w_a$ found throughout
Table~I: once the equation of state is allowed to evolve, the same data no longer need to
push its present-day value past $-1$.

\subsection{Goodness-of-fit and model selection}

For the full CC+BAO+Pantheon+CMB combination, $\chi^2_{\rm tot}$ decreases monotonically from $1560.81$ (wCDM) to $1552.64$ (CPL) to $1541.72$ (CPL+$\Sigma m_\nu$), driven mainly by the CC and Pantheon+ contributions, with BAO and CMB essentially unchanged across models — indicating the improvement is not an artifact of a single probe.

To test whether this improvement justifies the added complexity, we use
\begin{equation}
	\mathrm{AIC} = \chi^2_{\rm tot} + 2k, \qquad \mathrm{BIC} = \chi^2_{\rm tot} + k\ln N,
\end{equation}
with $k$ the number of free parameters. Table~\ref{tab:model_comparison_simple} is updated to reflect the correct parameter counts: CPL is fit with four parameters $(H_0,\Omega_m,w_0,w_a)$ and CPL+$\Sigma m_\nu$ with five $(H_0,\Omega_m,w_0,w_a,\Sigma m_\nu)$, consistent with Tables~\ref{tab:cpl_summary} and \ref{tab:cpl_mnu_summary}.

\begin{table}[H]
	\centering
	\caption{Comparison of wCDM, CPL, and CPL+$\Sigma m_\nu$ models, including the number of free parameters, total $\chi^2$, and information criteria. Sample size $N = N_{\rm CC}+N_{\rm BAO}+N_{\rm Pantheon+}+N_{\rm CMB} = 32+13+1701+3 = 1749$ (CMB contribution via the compressed distance-prior likelihood $\{R,\ell_a,\Omega_b h^2\}$), giving $\ln N \simeq 7.467$.}
	\label{tab:model_comparison_simple}
	\begin{tabular}{lccccc}
		\hline\hline
		Model & Free parameters & $N_{\rm par}$ & $\chi^2_{\rm tot}$ & AIC & BIC \\
		\hline
		wCDM & $(\Omega_m, H_0, \omega_{0})$ & 3 & $1560.81$ & $1566.81$ & $1583.21$ \\
		CPL & $(\Omega_m, H_0, w_0, w_a)$ & 4 & $1552.64$ & $1560.64$ & $1582.51$ \\
		CPL+$\Sigma m_\nu$ & $(\Omega_m, H_0, w_0, w_a, \Sigma m_\nu)$ & 5 & $1541.72$ & $1551.72$ & $1579.06$ \\
		\hline\hline
	\end{tabular}
\end{table}

\noindent
Even with a sample size as large as $N=1749$ — where the BIC penalty $k\ln N$ substantially exceeds the AIC penalty $2k$ — both criteria still favor the extended models: AIC decreases monotonically from $1566.81$ (wCDM) to $1560.64$ (CPL) to $1551.72$ (CPL+$\Sigma m_\nu$), and BIC follows the same ordering ($1583.21 \to 1582.51 \to 1579.06$). The BIC margin between models is noticeably smaller than the AIC margin, as expected from its stronger complexity penalty, yet CPL+$\Sigma m_\nu$ remains the preferred model despite carrying two additional parameters relative to wCDM. This confirms that the improvement in fit driven by dynamical dark energy and massive neutrinos is not an artifact of overfitting, but a genuine gain that survives even the more conservative BIC criterion.

\section{Comparison of Bayesian PINN with Full-Dataset MCMC}

The primary goal of this comparison is to assess the ability of a Bayesian physics--informed neural network (BPINN), trained primarily on late--time cosmological data, to reproduce posterior constraints on several parameters that are broadly comparable to those obtained from a full-dataset MCMC analysis. The comparison is intended solely at the level of parameter constraints and does not imply equivalence in methodology or dataset coverage. It should be noted that while the full MCMC analysis consistently models massive neutrinos at both the background and perturbation levels, the Bayesian PINN treats $\Sigma m_\nu$ as an effective late-time parameter entering only the background expansion. 
Therefore, the comparison focuses primarily on late-time geometric constraints rather than on a fully equivalent treatment of neutrino physics.

In addition to the datasets employed in the Bayesian PINN analysis (including CMB distance priors and late-time probes), we constructed a \emph{full dataset} for the MCMC analysis by incorporating the complete CMB information. Specifically, we used the latest large-scale CMB temperature and polarization angular power spectra from the final \emph{Planck 2018} data release, namely the \texttt{plikTTTEEE+lowl+lowE} likelihoods \cite{Aghanim}, and included the CMB lensing reconstruction power spectrum from the Planck 2018 trispectrum analysis \cite{Aghanim1}. These datasets were jointly analyzed within the \texttt{Cobaya} framework to perform a full MCMC exploration of the cosmological parameter space, providing a benchmark for comparison with the Bayesian PINN results.
To ensure a consistent and meaningful comparison between the BPINN and MCMC frameworks, we emphasize that the two approaches are not based on identical data configurations. While the MCMC analysis employs the full Planck CMB likelihood together with CMB lensing, BAO, CC, and Pantheon+ datasets, the BPINN framework utilizes CMB distance priors and excludes CMB lensing due to computational constraints associated with embedding full Boltzmann solvers. Therefore, the comparison should be interpreted as a consistency check rather than a strict likelihood-to-likelihood equivalence.

We find that BPINN can reproduce constraints on several late-time parameters, such as the Hubble constant $H_0$ and the dark energy equation-of-state parameters $(w_0, w_a)$, at a level broadly consistent with the full-dataset MCMC, typically within $1$--$2\sigma$. This demonstrates that BPINN provides an efficient and complementary framework for extracting a substantial fraction of the late-time geometric information, while being less sensitive to early-time modeling details and computationally less demanding. It should be emphasized, however, that BPINN is not a substitute for full Boltzmann-based MCMC pipelines, particularly for parameters primarily constrained by early-time physics.

\begin{table}[H]
	\centering
	\caption{Summary of CPL parameter constraints and Hubble tension levels with respect to Planck 2018 and SH0ES (R22) using full CMB dataset in MCMC method.  Uncertainties for cosmological parameters are at the $1\sigma$ level and for $\Sigma m_\nu$ [eV] at $2\sigma$ level.}
	\label{tab:cpl_h0_tension}
	\begin{tabular}{lccccc|cc}
		\hline
		Dataset & $\Sigma m_\nu$ [eV] & $\Omega_m$ & $w_0$ & $w_a$ & $H_0$ [km/s/Mpc] & $T_\text{Planck}$ & $T_\text{R22}$ \\
		\hline
		CMB+Pantheon+Lensing & $<0.21$ & $0.297 \pm 0.015$ & $-0.897 \pm 0.26$ & $-1.59 \pm 0.3$ & $72.15 \pm 1.9$ & 2.42 & 0.41  \\
		CMB+CC+Lensing      & $<0.2 $ & $0.300 \pm 0.016$ & $-1.01 \pm 0.31$ & $-0.729 \pm 0.33$ & $71.05 \pm 1.7$ & 2.06 & 1.00 \\
		CMB+BAO+Lensing     & $<0.19$ & $0.312 \pm 0.004$ & $-0.58 \pm 0.25$ & $-1.68 \pm 0.45$ & $68.65 \pm 1.5$ & 0.79 & 2.41 \\
		CMB+ALL             & $<0.14$ & $0.302 \pm 0.005$ & $-0.77 \pm 0.160$ & $-0.920 \pm 0.28$ & $70.12 \pm 1.3$ & 1.95 & 1.75 \\
		CMB+Lensing         & $<0.3$  & $0.318 \pm 0.020$ & $-1.043 \pm 0.070$ & $-1.41 \pm 0.35$ & $69.04 \pm 1.6$ & 0.98 & 2.10 \\
		\hline
	\end{tabular}
\end{table}
The results obtain for  $\Sigma m_\nu$ [eV], $\Omega_m$, $H_0$ [km/s/Mpc] for MCMC method are in fully agreement with \cite{Yarahmadi2025EPJC, Yarahmadi2025CTP, Yarahmadi2025IJMPA, Yarahmadi2025PDU, Yarahmadi2025AP, Yarahmadi2025JHEAp} and for  $w_0$,  $w_a$ in agreement with \cite{DESIDR2_2025,LiuLiWang2025,Malekjani2025,Park2024}.

\begin{figure}[H]
	\includegraphics[width=\columnwidth]{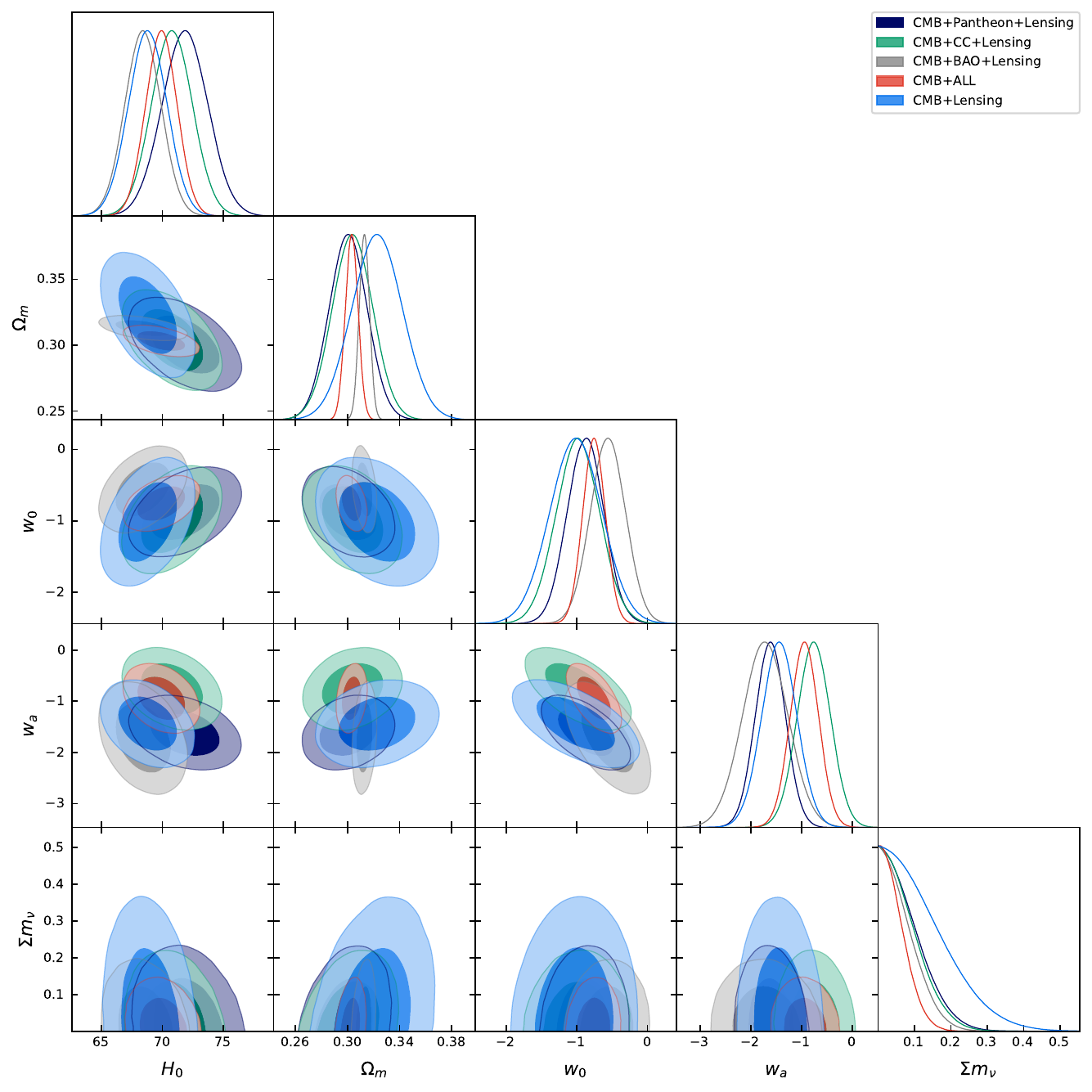}
	\caption{Corner plot of the CPL+$\Sigma m_\nu$ posterior distributions obtained from the full-dataset MCMC analysis for different combinations of CMB, BAO, Pantheon, and Cosmic Chronometer data, all including the complete Planck 2018 CMB likelihood and CMB lensing. The filled contours correspond to the $68\%$ and $95\%$ confidence levels. Each color represents a different dataset combination as indicated in the legend.}
	\label{fig:cpl9}
\end{figure}

Figure~\ref{fig:cpl9} displays the corner plot of the CPL+$\Sigma m_\nu$ posterior distributions for the various dataset combinations, derived from the full-dataset MCMC analysis with the complete Planck 2018 CMB likelihood and lensing. The contours correspond to the $68\%$ and $95\%$ confidence levels, illustrating the parameter correlations among $H_0$, $\Omega_m$, $w_0$, $w_a$, and $\Sigma m_\nu$.

The results presented in Tables~\ref{tab:cpl_h0_tension} and \ref{tab:cpl_mnu_summary} play a central role in assessing the consistency between different inference frameworks and the constraining power of the employed datasets. As shown in these tables, the analysis based on the \emph{full dataset} using a standard Markov Chain Monte Carlo (MCMC) approach implemented in the \texttt{Cobaya} framework—including CMB, BAO, Pantheon+, CC, and lensing data—yields constraints on the Hubble constant $H_0$ and the total neutrino mass $\Sigma m_\nu$. The Bayesian Physics-Informed Neural Network (PINN) approach, trained only on CMB distance priors combined with late-time cosmological probes, produces parameter regions that are broadly consistent with those obtained from the full MCMC analysis.

This agreement indicates that the Bayesian PINN can capture the main cosmological trends and reproduce the qualitative structure of the parameter space, despite using a reduced representation of the CMB information. In particular, the inferred ranges for $H_0$ and $\Sigma m_\nu$ overlap with the MCMC constraints within their uncertainties, which supports the PINN-based inference as a proof-of-concept. One reason this comparison is worth making explicit is that lensing is among the CMB
observables most directly tied to $\Sigma m_\nu$: neutrino free-streaming suppresses the
lensing potential on small scales, so a background-only fit that never sees lensing data
would be a natural place for a genuine bias in $\Sigma m_\nu$ to show up. Table~\ref{tab:bpinn_lensing}
lines up the four dataset combinations common to both pipelines. In every case the BPINN
and the MCMC+lensing values of $H_0$ agree to within $1\sigma$, and the $\Sigma m_\nu$
upper limits sit close together as well -- matching almost exactly for CMB+BAO and staying
within about $0.02$--$0.03\,\mathrm{eV}$ for the others. We do not want to read too much
into this: the MCMC runs also include the full Planck likelihood, not just lensing, so
this is not a controlled test that isolates the effect of lensing on its own. Still, a
purely background, late-time fit landing this close to a pipeline with direct access to
structure-growth information is a reasonable, if modest, indication that the geometric
data used here are not missing something essential about the neutrino sector, at least at
the precision current data allow.

\begin{table}[H]
	\centering
	\caption{Comparison of the BPINN (CMB distance priors only) and full-dataset MCMC (with CMB
		lensing) results for the four dataset combinations common to both pipelines.}
	\label{tab:bpinn_lensing}
	\begin{tabular}{lcccc}
		\hline\hline
		Dataset & $H_0^{\rm BPINN}$ & $H_0^{\rm MCMC+lens}$ & $\Sigma m_\nu^{\rm BPINN}$ & $\Sigma m_\nu^{\rm MCMC+lens}$ \\
		\hline
		CMB+CC             & $69.88\pm1.50$ & $71.05\pm1.7$ & $<0.26$ & $<0.20$ \\
		CMB+Pantheon       & $70.91\pm1.70$ & $72.15\pm1.9$ & $<0.28$ & $<0.21$ \\
		CMB+BAO            & $69.73\pm1.38$ & $68.65\pm1.5$ & $<0.19$ & $<0.19$ \\
		Full combination   & $70.41\pm1.28$ & $70.12\pm1.3$ & $<0.16$ & $<0.14$ \\
		\hline\hline
	\end{tabular}
\end{table}

Moreover, we observe that in some cases, the constraints on $H_0$ obtained with the Bayesian PINN are comparable to, or slightly narrower than, those derived from the full MCMC analysis. This behavior is likely driven by the implicit regularization induced by the physics-informed loss function and the reduced parameter degeneracies at late times, rather than by additional early-time information.

Importantly, this performance is achieved at a fraction of the computational cost: while the full MCMC analysis with \texttt{Cobaya} typically requires extensive computational resources, the Bayesian PINN approach converges within a significantly shorter runtime, often by orders of magnitude. This substantial gain in computational efficiency, combined with its ability to preserve the main cosmological trends, supports the Bayesian PINN as a promising and efficient tool for precision cosmology, particularly in studies of the Hubble tension and neutrino mass constraints.
\color{black}

Table~\ref{tab:pinn_vs_mcmc} summarizes the main differences and complementary features of the Bayesian PINN approach compared to a full-dataset MCMC analysis. While MCMC with the complete Planck 2018 CMB data, lensing, and late-time probes remains the standard for precision cosmology, the BPINN framework offers several advantages, including model flexibility, embedded physics constraints, efficient uncertainty quantification, and rapid inference once trained.  

\begin{table}[H]
	\centering
	\caption{Comparison between Bayesian PINN and full-dataset MCMC.}
	\label{tab:pinn_vs_mcmc}
	\begin{tabular}{|l|c|c|}
		\hline
		\textbf{Feature} & \textbf{Bayesian PINN} & \textbf{Full-Dataset MCMC} \\
		\hline
		Model Flexibility & High (implicit models) & Moderate (explicit models) \\
		Physics Constraints & Embedded in loss function & Hard-coded in solver \\
		Uncertainty Estimation & Dropout / Variational methods & Posterior sampling \\
		Speed (after training) & Fast inference & Slower due to sampling \\
		Data Integration & Direct, no separate likelihood & Requires likelihood functions \\
		Requires Solver & No (differentiable loss) & Yes (CAMB or CLASS) \\
		Ease of Modifying Physics & Easy & Requires code changes \\
		Scalability & High (parallelizable) & Limited by sampler speed \\
		\hline
	\end{tabular}
\end{table}

\section{Conclusion}
\label{sec:conclusion}

In this work we have presented a late-time cosmological analysis of
the Chevallier--Polarski--Linder (CPL) dark energy parametrization, in
both its bare form and its extension with a free summed neutrino mass
$\Sigma m_\nu$, within a fully Bayesian Physics-Informed Neural
Network (BPINN) framework. By embedding the background Friedmann
equation directly into the network loss, our approach reconstructs
the late-time expansion history $H(z)$ in a manner that is
simultaneously data-driven and physically consistent, while a
dropout-based variational scheme propagates epistemic uncertainty
through the resulting posteriors. The analysis combines Cosmic
Chronometers, the Pantheon+ Type Ia supernova sample, and DESI DR2
baryon acoustic oscillations, anchored throughout by Planck 2018 CMB
distance priors, allowing us to examine the joint role of dark energy
dynamics and neutrino mass in the Hubble tension within a single,
internally consistent pipeline.

Across every dataset combination considered, the reconstructed CPL
parameters remain close to, yet measurably distinct from, the
$\Lambda$CDM limit. The present-day equation of state is consistently
found in the quintessence regime, $w_0 \gtrsim -1$, with best-fit
values ranging from $-0.96$ to $-0.55$, while the evolution parameter
$w_a$ is consistently negative, typically between $-1.7$ and $-0.6$.
This combination describes an equation of state that becomes mildly
more negative at higher redshift without crossing the phantom divide
at the present epoch. Because the associated uncertainties remain
sizable, a cosmological constant cannot be excluded at the $1\sigma$
level for any dataset combination; nevertheless, the systematic
upward shift in $H_0$ produced by the CPL extension -- most
pronounced for combinations that include Pantheon+ -- translates into
a substantial relief of the tension with the SH0ES (R22) measurement,
falling to $0.56\sigma$ for CPL (CMB+Pantheon) and to $0.83\sigma$ for
CPL+$\Sigma m_\nu$ (CMB+CC+Pantheon).

The inclusion of a free neutrino mass does not alter this qualitative
picture, but it stabilizes the inferred parameters within a narrower
range, $H_0 = 69.7$--$71.6~\mathrm{km\,s^{-1}\,Mpc^{-1}}$ and
$\Sigma m_\nu \lesssim 0.16$--$0.28~\mathrm{eV}$ at $1\sigma$,
tightening progressively as complementary probes are added. For the
full CMB+CC+BAO+Pantheon combination, we obtain
$\Sigma m_\nu < 0.16~\mathrm{eV}$ together with a markedly asymmetric
resolution of the Hubble tension, $T_{\rm R22} = 1.51\sigma$ against
$T_{\rm Planck} = 2.5\sigma$. This asymmetry -- a pronounced relief of
the local tension accompanied by a persistent, more moderate
disagreement with Planck -- recurs across all dataset combinations,
and indicates that late-time modifications of this kind redistribute,
rather than eliminate, the discrepancy between early- and
late-Universe determinations of $H_0$.

The preference for the extended models is not an artifact of their
additional flexibility. The total $\chi^2$ decreases monotonically
from wCDM to CPL to CPL+$\Sigma m_\nu$, and both the Akaike
and, more conservatively, the Bayesian information criteria favor
CPL+$\Sigma m_\nu$ despite its two extra parameters relative to
wCDM. As an independent check, we compared the BPINN
posteriors against a full-dataset MCMC analysis employing the
complete Planck 2018 \texttt{plikTTTEEE+lowl+lowE} likelihood together
with CMB lensing, implemented in the \textsc{Cobaya} framework. The
two approaches agree on $H_0$, $(w_0, w_a)$, and $\Sigma m_\nu$
typically within $1$--$2\sigma$, even though the BPINN relies only on
compressed CMB distance priors and treats $\Sigma m_\nu$ purely at
the background level. This is particularly relevant for $\Sigma m_\nu$: lensing is one of the observables most
sensitive to neutrino mass, and yet the background-only BPINN fit lands close to the
lensing-informed MCMC result for every combination we could compare, suggesting that the
late-time geometric information used here already captures much of what matters for this
parameter, at least at current precision. This agreement shows that the Bayesian PINN
captures the dominant late-time cosmological trends at a small
fraction of the computational cost of a full Boltzmann-based MCMC
run, without introducing a significant bias relative to the standard
pipeline.

Two caveats are worth stressing. First, the inferred nature of dark
energy is sensitive to which probes are combined: DESI DR2 BAO data
alone permit, at the $2\sigma$ level, a region of parameter space
extending into the phantom regime, but this preference is washed out
once Cosmic Chronometers, Pantheon+, and CMB distance priors are
folded in, all of which pull the posterior back toward $w_0 > -1$.
Second, as implemented here $\Sigma m_\nu$ enters only through the
late-time background expansion; the resulting bounds should therefore
be read as effective, late-time constraints rather than as a
substitute for a full perturbative treatment of massive neutrinos.

Taken together, our results support a coherent, if incomplete,
picture in which a mildly evolving, quintessence-like dark energy
component, together with a sub-eV neutrino mass, meaningfully
alleviates the tension between cosmological and local determinations
of $H_0$ while leaving a moderate residual disagreement with Planck.
Methodologically, this work supports the use of the Bayesian PINN as
an efficient complement to standard MCMC pipelines for late-time
cosmology, particularly for models with strong internal parameter
degeneracies. Extending this framework to a perturbative
treatment of massive neutrinos, and confronting it with
next-generation BAO, supernova, and structure-growth data, will help
determine whether the residual Hubble tension signals genuine physics
beyond $\Lambda$CDM or remains within reach of late-time
phenomenological extensions.

\section*{Acknowledgments}
We would also like to extend our special thanks to Prof. S. D. Odintsov for his valuable
comments and insightful discussions, which helped us improve the physical interpretation
of the dark-energy sector presented in this work.
This work is based upon research funded by Iran National Science Foundation 
(INSF) under project No.4036326

\end{document}